\documentclass[12pt]{iopart}

\usepackage{graphicx}
\usepackage{amsmath,amssymb}
\usepackage{dcolumn}
\usepackage{bm}
\usepackage{dsfont}
\usepackage{color}
\usepackage{amsthm}
\definecolor{labelkey}{cmyk}{.4,.2,0,0}
\usepackage{appendix}
\usepackage{tocloft}

\newcommand{\sgn}{\textup{sgn}}
\newcommand{\expval}[1]{{\left\langle #1 \right\rangle}}
\newcommand{\diff}{\mathrm{d}}

\newtheorem{thm}{Theorem}[section]

\newtheorem{lem}[thm]{Lemma}
\newtheorem{prop}[thm]{Proposition}

\newtheorem{defn}[thm]{Definition}

\begin{document}

\title[Eigenvalues \& singular values of low-rank NNEMs]{Statistics of the non-zero eigenvalues and singular values of low-rank random matrices with non-negative entries}

\author{Mark J. Crumpton, Yan V. Fyodorov and Pierpaolo Vivo}

\address{Department of Mathematics, King's College London, London WC2R 2LS, United Kingdom}
\ead{pierpaolo.vivo@kcl.ac.uk}
\vspace{10pt}
\begin{indented}
\item[]January 2025 
\end{indented}

\begin{abstract}
We compute analytically the probability distribution and moments of the sum and product of the non-zero eigenvalues and singular values of random matrices with (i) non-negative entries, (ii) fixed rank, and (iii) prescribed sums of the entries in each row. Applications of such matrices are discussed in the context of Markov chains, economics and social networks to name a few. All results are valid at finite matrix size and are given in terms of the statistics of vectors of general Dirichlet random variables. Analytical results are corroborated by numerical simulations throughout with excellent agreement.
\end{abstract}

\tableofcontents

\section{Introduction}

Matrices with non-negative entries (NNEMs) and prescribed row sums occur in many applications. For instance, they prominently feature in the context of Markov chain dynamics \cite{MarkovTextBook, Seneta1981}, describing random walks on complex networks. In such dynamics, one usually starts with assigning the entries $A_{ij}$ of the adjacency matrix $A$ the values $\{0,1\}$, depending on whether an edge exists between the vertices $i$ and $j$ of a graph representing the network. The adjacency matrix is then normalised according to the out-degree $k_i$ of each vertex, to give a normalised transition-probability matrix $\pi_{ij}=A_{ij}/k_i$ \cite{Masuda2017}, representing the (in this simple setting, unbiased) probability that the walker reaches vertex $j$ starting from vertex $i$ in one step of the process. Such matrices whose $(i,j)$ entry is the probability that a Markov process jumps from site $i$ to site $j$ are known as stochastic, and their properties determine the long-time equilibrium distribution of the associated random walker \cite{Hendricks1972}. Analytical and numerical results are available for row (column) stochastic matrices, where the entries in each row (column) sum to one, and doubly stochastic matrices, where the entries in all rows and columns sum to one \cite{Johnson1981, Sinkhorn1967}. In particular, conditions for similarity between row and doubly stochastic matrices, limits of convergence of a non-negative matrix to a doubly stochastic matrix by alternate scaling of rows and columns, mixing time of Markov chains, methods to generate doubly stochastic matrices and limiting spectral density of row stochastic matrices all have been intensively studied.

Another prominent application of NNEMs with fixed row sums (which arise as global economic constraints called accounting identities) comes in the field of economics, namely the input-output matrices developed by Leontief \cite{Leontief1970, BCCV23, BCCV24}. These matrices are used to assess how much of the output of one economic sector is needed as an input to another sector for a complex economy to work, providing quantitative indicators (such as \emph{upstreamness} and \emph{downstreamness} \cite{McNerney2021, Miller2017, Antras2012}) of relationships and inter-dependence between different sectors \cite{MillerBlair2009}. The study of NNEMs arises naturally in a wide range of economic areas, due to the non-negative nature of many economic quantities, such as exchange rates, expenses, and productivity measures \cite{Bray1922, Ivanov2001, Zeng2008, Debreu1953}. Additionally, recent research into the so-called Leontief inverse, which relates total output and demand of economic sectors, has led to the realisation that these matrices have universal features that can be applied to a wide range of economies \cite{VivoInversionFree2020}.

A further application of fixed row sum NNEMs comes in the study of opinions in social networks, known broadly as DeGroot learning. DeGroot learning considers the evolution of the opinions of a group of participating agents on a particular topic \cite{Golub2010}. Each agent has an initial opinion on the subject and also a weighted opinion of all other agents, quantifying how much they trust the other agents. The weighted opinions form a stochastic `trust' matrix and so the process of the evolution of the opinions on the associated topic can be modelled as a Markov chain. In the process of opinion formation  the opinions reach a stationary state after a large number of interactions, providing a useful model helping to address consensus formation in complex societies \cite{Sikder2020, FJ1999, BCN2020}. Additional applications of NNEMs can be found in Game Theory, for example the payoff matrix of non-zero games, which is used to determine optimal in-game strategies \cite{Raghavan1965}. Many more interesting problems that rely on the study of NNEMs are outlined in \cite{Bapat1997}. Among a few examples worth mentioning are the pricing of houses and optimal scheduling of tasks in work environments with the aim of creating fair divisions of quantities, while respecting the needs of individuals \cite{Bai2022, Amanatadis2022}.

The central result about NNEMs spectra is the famous Perron-Frobenius theorem telling that every NNEM will have one real eigenvalue $\lambda_r$, such that all the other (in general complex) eigenvalues have magnitudes $\leq |\lambda_r|$. Additionally, it states that the eigenvector corresponding to $\lambda_r$ will have strictly non-negative elements, while all other eigenvectors must have at least one negative element. Among other things, the magnitude of the largest (Perron-Frobenius) eigenvalue and its distance from the second-largest eigenvalue is known to govern various phenomena, including many features of dynamical processes on networks \cite{Pillai2005}. For example, in the context of the mean first passage time that a random walker takes to reach a target node on a complex network, a large gap between the Perron-Frobenius eigenvalue of the transition matrix and the bulk of all the other eigenvalues guarantees that a fast, approximate formula works without the need for heavy matrix inversions \cite{BCCV21, FGTVA22, KKRA2020}.

From that perspective one can be interested in  studying statistical properties of the Perron-Frobenius and the next largest eigenvalues in ensembles of NNEMs with random entries, hoping to reveal generic and potentially universal spectral features shared by NNEMs of diverse nature.
As one of the simplest examples, one may consider Gaussian perturbations of rank-$1$ matrices. The ensuing interplay between the bulk eigenvalues (governed by the Girko-Ginibre circular law \cite{Ginibre1965, Girko1985}) and potential outliers has profound applications in data processing \cite{Benaych-Georges2011, Tao2013}. Extensions involving positivity constraints on the rank-$1$ matrix reveal connections to Markov chains, where a temperature-like parameter controls a phase transition so that below a certain temperature the standard circular law fails, leading to  eigenvalue spectra of different type \cite{deGiuli2021}. 

While the study of Extreme Value Statistics is probably one of the most developed corners of Random Matrix Theory since the discovery of the celebrated Tracy-Widom distribution \cite{TracyWidom1994}, the associated research on random NNEMs is rather limited \cite{Chi, Bordenave, Coste}. Instead, the main effort has been mostly focused either on the  characterization of eigenvalues or singular values of random NNEMs in the bulk of the spectrum
\cite{Chatterjee2010,Cappellini2009, Chafai2010,Bordenave2012, Nguyen2014} or on properties of the associated permanents, averages, and products \cite{Jerrum2004, Dhara2020, Hognas2005, Hennion1997, Mierczynski2015,Innocentini2018}. In fact extreme eigenvalues (or singular values) of generic NNEMs seem to be quite difficult to study, and they have been  mainly addressed via bounds, iterations, and inequalities \cite{Restrepo2007,Ng2009,Nikiforov2002}. In particular, in the random setting one may notice that the non-negativity constraint on the entries is bound to break the rotational invariance of any model - thus ruling out the most powerful analytical tools available (orthogonal polynomials, free probability, determinantal processes etc.). Furthermore, it also non-trivially restricts the domain of integration in relevant random variables, and in this way forces non-Gaussian multiple integrals into the standard cavity or replica treatments \cite{KTW10, SVK19, SVK20, SVK21} - which quickly leads to insurmountable computational challenges.

From this angle, one way to make progress in the analytical study of statistics of extreme eigenvalues of random NNEMs with fixed row sums is to consider a simplified scenario where the matrix has a low-rank. Low-rank matrix approximations are known to be very useful, as, in this case, there are only a handful of non-zero eigenvalues/singular values that need to be taken into consideration, which can usually be easily linked to the row/column sums \cite{TAD22}. In fact, thanks to the Eckart-Young-Minsky theorem, it is known that if one has full knowledge of all spectral properties of a matrix then the best specified rank approximation is obtained using a singular value decomposition \cite{Stewart1993}. Other low-rank approximations, specific to NNEMs, have also been considered, for example \cite{Cohen1993}. Moreover, such approximations also arise naturally in several `matrix reconstruction' schemes based on the Maximum Entropy method, with applications e.g. to inter-bank network reconstruction \cite{Squartini2017, Squartini2018, Cimini2015}. Approximations using low-rank matrices have also been utilised in the study of dissipative quantum transport, the $N$-body problem and fluid dynamics \cite{Zeng2013, Yokota2017, Lestandi2021}.

In this work, we analytically characterise statistical features of the non-zero eigenvalues and singular values of random fixed rank NNEMs with prescribed row sums. All theoretical results are validated through numerical simulations with excellent agreement. The models studied in this work are rare examples of $N\times N$ non-invariant matrices where the full Extreme Value Statistics and certain spectral observables can be computed exactly \emph{for finite $N$} (and not just for large $N$). The remainder of this manuscript is arranged as follows. In Section \ref{sec:Defs}, we present models for general rank-$1$, rank-$2$ and rank-$R$ NNEMs with fixed row sums, which are defined in terms of Dirichlet random vectors. Then, Sections \ref{sec:Res_M1}, \ref{sec:Res_M2} and \ref{sec:Res_MR} are devoted to presenting results for the statistical properties (i.e. probability distributions and integer moments) of the sum and product of the non-zero eigenvalues and singular values of our rank-$1$, rank-$2$ and rank-$R$ NNEMs respectively. We then provide a brief discussion of our results and suggest some open problems in Section \ref{sec:Discussion}. Our results are regularly verified through the use of numerical simulations and are then proved in detail in the following sections. Specifically, in Sections \ref{sec:rank1_proofs}, \ref{sec:rank2_proofs} and \ref{sec:rankR_proofs} we prove the results for our random rank-$1$, rank-$2$ and rank-$R$ models respectively. Finally, in \ref{app:sym_funcs_rankR} we verify a useful identity relating to the equivalence of expressions given in terms of symmetric functions. This identity helps us to derive results for the distribution of the sum of the eigenvalues in the three different models considered in this work.

\section{Definition of random matrix models and statement of main results} \label{sec:Main}

\subsection{Definition of models}
\label{sec:Defs}

The randomness associated with all $N \times N$ matrices studied in this work arises solely from sets of $N$ Dirichlet random variables (DRVs), which is a natural choice for non-negative random variables constrained to have a linear sum equal to unity. Associating DRV's with entries of a vector $\bm a\in \mathbb{R}_+^N$, their joint probability density function (jpdf) is chosen in the following form: 
\begin{equation}
    \mathcal{P}_{\bm k} (\bm a) = \frac{1}{B(\bm k)} \left[ \prod_{i=1}^{N} a_i^{k_i - 1} \right] \delta \left( \sum_{i=1}^N a_i - 1 \right) \hspace{1cm} \text{where} \hspace{1cm} B(\bm k) = \frac{\prod\limits_{i=1}^N \Gamma(k_i)}{ \Gamma(\kappa) } \ ,
    \label{eq:jpdf_GDRV}
\end{equation}
such that $\bm k$ is a general vector of size-$N$ containing positive integer entries, $k_i \geq 1$, and $\kappa = \sum_i k_i$. These will be referred to as the Dirichlet parameters. Furthermore, we have also made use of the Dirac delta-function, $\delta(\cdot)$. Note that there is an easy way to randomly sample the elements of a set of DRVs  by first generating a set of $N$ independent random variables $\{\gamma_i\}$, such that each element is drawn from a distribution 
\begin{equation}
    p_{k_i}(\gamma_i) = \frac{\gamma_i^{k_i-1} e^{-\gamma_i}}{\Gamma(k_i)} \Theta[\gamma_i]\ ,
\end{equation}
where $\Theta[x]$ is the usual Heaviside $\Theta$-function (such that $\Theta[x>0] =0$ and $\Theta[x\leq 0] = 0$). Then, the set of DRVs $\{a_i\}$ is generated by simply normalising these random variables as $a_i = \gamma_i/(\sum_j \gamma_j)$. Natural yet useful particular instances of Dirichlet Random Variables are the cases of (i) symmetric DRV's (corresponding to choosing all $k_i$ being equal) and (ii) flat DRV's (when further $k_i=1\, \forall i=1, \ldots, N$). The study of DRVs is a well-established area in probability and statistics, with applications in diverse fields ranging from Biology to Baseball \cite{Lange95, Null09}. 

Recall that a square matrix $M$ of size $N$ and rank $R\leq N$ is bound to have at most $R$ non-zero eigenvalues and singular values, the latter being defined as the square roots of the eigenvalues of $M^T M$. In what follows, we construct matrices of a specified rank $R$, by multiplying each element of a set of $R$ fixed $N$-vectors by the transpose of an independent Dirichlet random vector of the same size, then summing the resulting matrices. Generating matrices in this way yields matrices with fixed row sums (that can be, in principle, prescribed according to some application-motivated set of row sums).  For the remainder of this work we will mostly consider the properties of a rank-$1$ and a rank-$2$ NNEM in detail, and then eventually provide some statements and results that apply to NNEMs of an arbitrary fixed rank-$R$. Specifically, we will present results pertaining to the sum and product of the eigenvalues and singular values of the specified rank NNEMs. Some of these results will be stated in the form of probability density functions (pdfs) denoted as $\mathcal{P}$, whilst others are specified in terms of moments, with averages denoted as $\expval{\cdot}_{E}$, the subscript $E$ standing for the ensemble we are averaging over. 

We now define the NNEMs that will be studied throughout this work, along with equations for the sum and product of the non-zero eigenvalues and singular values in each model \footnote{It is useful to remember that for an $N\times N$ matrix $AB$ of rank $R \leq N$, the $R$ non-zero eigenvalues of $AB$ are equal to the non-zero eigenvalues of $BA$, where $A$ and $B$ are matrices of size $N \times R$ and $R \times N$ respectively \cite{Nakatsukasa2019}.}.

\begin{defn}\label{def:M1}
    Let $M_1$ be a generic $N\times N$ rank-$1$ NNEM constructed as
    \begin{equation}
        M_1  \equiv \bm z \bm a^T= \left( \begin{matrix} 
        a_1 z_1 & a_2 z_1 & \cdots & a_N z_1 \\
        a_1 z_2 & a_2 z_2 & \cdots & a_N z_2 \\
        \vdots & \vdots & \ddots & \vdots  \\
        a_1 z_N & a_2 z_N & \cdots & a_N z_N
        \end{matrix} \right)\ .
        \label{eq:M1_def}
    \end{equation}
    In the above $\bm a$ is a general Dirichlet random vector with Dirichlet parameters arranged as the entries of the vector $\bm k$ and their linear sum being denoted as $\kappa = \sum_i k_i$. Further to this, $\bm z$ is a fixed non-negative vector, such that the $i$th entry of $\bm z$ corresponds to the sum of the entries in row $i$ of $M_1$. The matrix $M_1$ will have one non-zero eigenvalue and one non-zero squared singular value given by
    \begin{equation}
        \lambda = \bm a^T \bm z \hspace{1cm} \text{and} \hspace{1cm} \sigma^2 =  r^2 \varphi \ ,
        \label{eq:lambda_svs_def} 
    \end{equation}
    respectively, where $r^2 \equiv \bm z^T \bm z$ and $\varphi \equiv \bm a^T \bm a$.
\end{defn}

\begin{defn} \label{def:M2}
Let $M_2$ be a generic $N \times N$ rank-2 NNEM constructed as
\begin{equation}
    M_2 \equiv \bm x \bm a^T + \bm y \bm b^T \ ,
    \label{eq:M2_def}
\end{equation} 
where $\bm a$ and $\bm b$ are two independent Dirichlet random vectors of length $N$, sampled from the jpdf in Eq. \eqref{eq:jpdf_GDRV} with Dirichlet parameters given by the entries of the vectors $\bm h$ and $\bm k$ respectively, such that $\eta = \sum_i h_i$ and $\kappa = \sum_i k_i$. Furthermore, $\bm x$ and $\bm y$ are fixed non-negative vectors such that the sum of the entries in row $i$ of $M_2$ is given by $x_i + y_i$ for $i = 1,2,\ldots,N$. Such a matrix $M_2$ will have two non-zero eigenvalues given by
\begin{equation}
    \lambda_{1,2} = \frac{T}{2} \pm \frac{1}{2} \sqrt{T^2 - 4D} \ ,
\end{equation}   
where the sum and product of the non-zero eigenvalues can be written as
\begin{align}
    T \equiv& \lambda_1 + \lambda_2 = \bm a^T \bm x + \bm b^T \bm y \label{eq:T_def} \equiv \alpha + \zeta \ , \\
    D \equiv& \lambda_1 \lambda_2 = (\bm a^T \bm x ) (\bm b^T \bm y ) - (\bm a^T \bm y ) (\bm b^T \bm x ) \equiv \alpha \zeta - \beta \gamma \ ,
    \label{eq:D_def} 
\end{align}
respectively. Similarly, $M_2$ will possess two non-zero singular values $ \sigma_{1}$ and $ \sigma_{2}$ whose squares are given by
\begin{equation}
    \sigma_{1,2}^2 = \frac{\tau}{2} \pm \frac{1}{2} \sqrt{\tau^2 - 4\Delta}\ ,
\end{equation}
such that their sum and product can be represented as
\begin{align}
    &\tau \equiv \, \sigma_1^2 + \sigma_2^2 = (\bm x^T \bm x) (\bm a^T \bm a) + (\bm y^T \bm y) (\bm b^T \bm b) + 2 (\bm x^T \bm y ) (\bm a^T \bm b) \label{eq:tau_def} \ , \\
    &\Delta \equiv \, \sigma_1^2 \sigma_2^2 = \Big( (\bm a^T \bm a) (\bm b^T \bm b) - (\bm a^T \bm b)^2 \Big) \Big( (\bm x^T \bm x) (\bm y^T \bm y) - (\bm x^T \bm y)^2 \Big) \label{eq:Delta_def} \ ,
\end{align}
respectively. 
\end{defn}

\begin{defn}\label{def:MR}
    Let $M_R$ be a generic rank-$R$ matrix of size $N \times N$ defined according to 
    \begin{equation}
        M_R \equiv \sum_{\ell=1}^R \bm z_\ell \bm a_\ell^T = Z^T A \ .
        \label{eq:def_MR}
    \end{equation}
    Here, for each $\ell = 1, 2, \ldots,R$, $\bm a_\ell$ is a general Dirichlet random vector, with Dirichlet parameters arranged in the vectors $\bm k_\ell$ and their linear sums being denoted as $\kappa_\ell = \sum_j(\bm k_\ell)_j$. Additionally, $\bm z_\ell$ is a fixed non-negative vector, such that the set of $\{\bm z_\ell\}$ govern the row sums of $M_R$, i.e. the elements in row $i$ of $M_R$ sum to $\sum_{\ell=1}^R (\bm z_\ell)_i$. Furthermore, we use the three sets of vectors $\{\bm z_\ell\}$, $\{\bm a_\ell\}$ and $\{\bm k_\ell\}$ to introduce three $R \times N$ matrices $Z, A$ and $K$ defined according to  
    \begin{equation}
        Z \equiv \left( \begin{matrix}
            \longleftarrow & \bm z_1^T & \longrightarrow \\
            \longleftarrow & \bm z_2^T & \longrightarrow \\
             & \vdots &  \\
            \longleftarrow & \bm z_R^T & \longrightarrow
        \end{matrix} \right) \, , \hspace{0.3cm} 
        A \equiv \left( \begin{matrix}
            \longleftarrow & \bm a_1^T & \longrightarrow \\
            \longleftarrow & \bm a_2^T & \longrightarrow \\
             & \vdots &  \\
            \longleftarrow & \bm a_R^T & \longrightarrow
        \end{matrix} \right)\, ,  \hspace{0.3cm} 
        K \equiv \left( \begin{matrix}
            \longleftarrow & \bm k_1^T & \longrightarrow \\
            \longleftarrow & \bm k_2^T & \longrightarrow \\
             & \vdots &  \\
            \longleftarrow & \bm k_R^T & \longrightarrow
        \end{matrix} \right)  \ ,
        \label{eq:ZKA_def}
    \end{equation}
    with entries $Z_{ij} = (\bm z_i)_j$ for $i = 1, 2, \ldots ,R$ and $ j = 1, 2, \ldots , N$, and analogously for the entries of $A$ and $K$. The important spectral information about the $R$ non-zero eigenvalues of $M_R$, $\lambda_i$ for $i = 1,2,\ldots,R$, can be obtained from the matrix
    \begin{equation}
        \Lambda_R \equiv A Z^T = \left( \begin{matrix}
            \bm a_1^T \bm z_1 & \bm a_1^T \bm z_2 & \cdots & \bm a_1^T \bm z_R  \\
            \bm a_2^T \bm z_1 & \bm a_2^T \bm z_2 & \cdots & \bm a_2^T \bm z_R  \\
            \vdots & \vdots & \ddots & \vdots \\
            \bm a_R^T \bm z_1 & \bm a_R^T \bm z_2 & \cdots & \bm a_R^T \bm z_R  \\
        \end{matrix} \right) \ ,
        \label{eq:Lambda_R_def}
    \end{equation}
    with the sum and product of the eigenvalues given by 
    \begin{align}
        T_R \equiv \Tr(M_R) = \Tr(\Lambda_R) = \sum_{i=1}^R \bm a_i^T  \bm z_i 
        \hspace{0.5cm} \text{and} \hspace{0.5cm}
        D_R \equiv \prod_{i=1}^R \lambda_i = \det(\Lambda_R) \ .
        \label{eq:def_T_D_MR}
    \end{align}
    Furthermore, the important information regarding the $R$ non-zero squared singular values of $M_R$, $\sigma_i^2$ for $i=1,2, \ldots , R$, can be obtained from the matrix
    \begin{equation}
        \Sigma_R \equiv AA^T ZZ^T = \left( \begin{matrix}
            \bm a_1^T \bm a_1 & \bm a_1^T \bm a_2 & \cdots & \bm a_1^T \bm a_R  \\
            \bm a_2^T \bm a_1 & \bm a_2^T \bm a_2 & \cdots & \bm a_2^T \bm a_R  \\
            \vdots & \vdots & \ddots & \vdots \\
            \bm a_R^T \bm a_1 & \bm a_R^T \bm a_2 & \cdots & \bm a_R^T \bm a_R  \\
        \end{matrix} \right) 
        \left( \begin{matrix}
            \bm z_1^T \bm z_1 & \bm z_1^T \bm z_2 & \cdots & \bm z_1^T \bm z_R  \\
            \bm z_2^T \bm z_1 & \bm z_2^T \bm z_2 & \cdots & \bm z_2^T \bm z_R  \\
            \vdots & \vdots & \ddots & \vdots \\
            \bm z_R^T \bm z_1 & \bm z_R^T \bm z_2 & \cdots & \bm z_R^T \bm z_R  \\
        \end{matrix} \right) \ .
    \end{equation}
    It is therefore natural to define the matrices $\Sigma_A \equiv AA^T$ with entries given by $(\Sigma_A)_{ij} = \bm a_i^T \bm a_j$, and analogously for $\Sigma_Z \equiv ZZ^T$. Within this notation, the sum and product of the $R$ non-zero squared singular values of $M_R$ are given by 
    \begin{align}
        \tau_R \equiv& \Tr(M_R^T M_R) = \Tr(\Sigma_R) = \sum_{i,j=1}^R (\bm a_i^T \bm a_j) (\bm z_j^T \bm z_i) \ , \label{eq:def_tau_MR} \\
        \Delta_R \equiv& \prod_{i=1}^R \sigma_i^2 = \det(\Sigma_R) = \det(\Sigma_A) \det(\Sigma_Z) \ ,
        \label{eq:def_Delta_MR}
    \end{align}
    respectively.
\end{defn}

In the following three subsections, we now present a summary of our results for various statistics related to the sum and product of the non-zero eigenvalues and squared singular values for each of the above NNEMs. Proofs of these results can be found in Sections \ref{sec:rank1_proofs}, \ref{sec:rank2_proofs}  and \ref{sec:rankR_proofs} for rank-$1$, rank-$2$ and rank-$R$ NNEMs respectively. The results presented for each rank can be easily programmed into symbolic software (i.e. Mathematica) - as we have done to generate all Figures that follow.

\subsection{Results for rank-1 NNEMs}
\label{sec:Res_M1}

In this section, we present our results pertaining to the statistical properties of the only non-zero eigenvalue and squared singular value of the rank-$1$ NNEM $M_1$, see Definition \ref{def:M1}. We start by providing the statistics of the non-zero eigenvalue below.

\begin{thm} \label{thm:lam_M1}
    Let $M_1 = \bm z \bm a^T$ be a random rank-$1$ $N\times N$ NNEM drawn according to Definition \ref{def:M1}. The pdf of the non-zero eigenvalue of $M_1$, $\lambda = \bm a^T \bm z$, reads
    \begin{equation}
        \mathcal{P}_{\bm z, \bm k}(\lambda) \equiv \bigg\langle \delta\left( \lambda - \bm a^T \bm z \right) \bigg\rangle_{\bm a} = \frac{\Gamma(\kappa)}{\Gamma(\kappa - 1)} \, \bm \partial_{\bm x}^{\bm k} \, \left[ \sum\limits_{j=1}^{N} \frac{\left(  x_j - \lambda \right)^{\kappa-2}}{\prod\limits_{l\neq j}^N (x_j - x_l)}  \Theta\left[ x_j - \lambda \right] \right]_{\bm x = \bm z}\ , 
        \label{eq:pdf_lambda_GDRV}
    \end{equation}
    where we have introduced the differential operator
    \begin{equation}
        \bm \partial_{\bm x}^{\bm k} = \prod_{i=1}^N \frac{1}{\Gamma(k_i)} \frac{\partial^{k_i-1}}{\partial x_i^{k_i-1}} \ .
        \label{eq:def_partials}
    \end{equation}
    One can verify this pdf to be non-vanishing only for $\lambda \in [\min(\bm z), \max(\bm z)]$. Correspondingly the non-negative integer moments ($m \geq 0$) of $\lambda$ can be derived from the above result as
    \begin{equation}
        \expval{\lambda^m}_{\bm a} \equiv \int_0^\infty \diff \lambda \, \lambda^m \, \mathcal{P}_{\bm z, \bm k}(\lambda) = \frac{\Gamma(\kappa) \, \Gamma(m + 1)}{\Gamma(\kappa + m)} \, \bm \partial_{\bm x}^{\bm k} \, \left[ \sum_{j=1}^N \frac{x_j^{\kappa + m -1}}{\prod\limits_{\ell \neq j}^N (x_j - x_\ell)} \right]_{\bm x = \bm z} \ .
        \label{eq:moms_lambda_GDRV}
    \end{equation}
\end{thm} 

\begin{figure}[h]
    \centering
    \includegraphics[scale = 0.3]{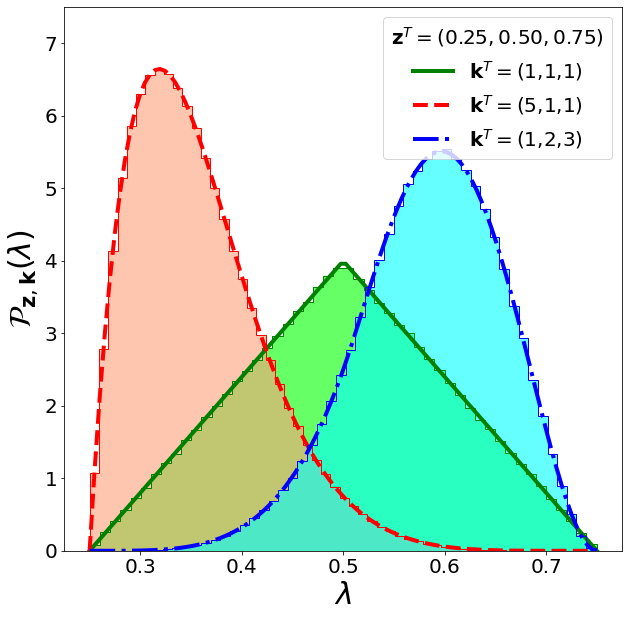}
    \hspace{1cm}
    \includegraphics[scale = 0.3]{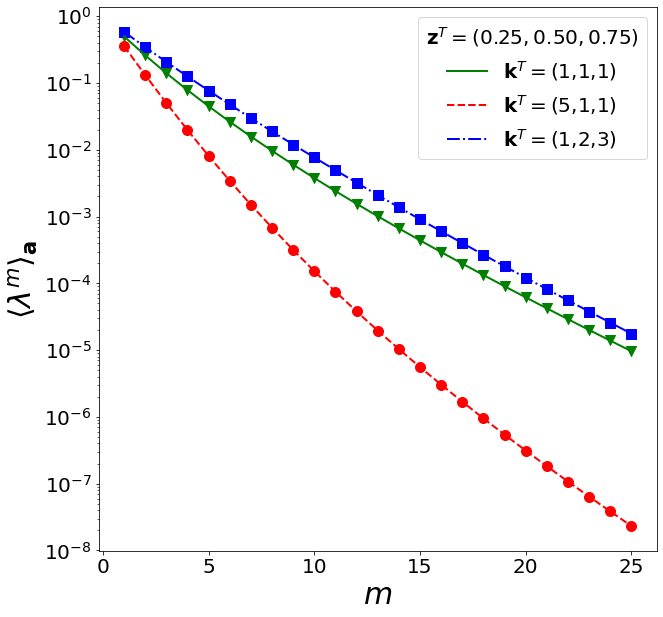}
    \caption{\small Statistics of the non-zero eigenvalue, $\lambda$, of the $3 \times 3$ rank-$1$ NNEM $M_1 = \bm z \bm a^T$ when $\bm a$ is drawn from a general Dirichlet distribution and $\bm z = (0.25, 0.50, 0.75)^T$. Left: Distribution of $\lambda$ according to Eq. \eqref{eq:pdf_lambda_GDRV} (coloured lines) compared to histograms of numerical data. Right: non-negative integer moments of $\lambda$ (coloured lines plotted point-to-point using Eq. \eqref{eq:moms_lambda_GDRV}) compared to numerically observed averages (coloured markers). Each of the data sets contains $10^6$ numerically generated samples of $\lambda$.}
    \label{fig:lam_M1}
\end{figure}

\noindent
The proof of these results is presented in Section \ref{sec:proofs_lam_M1}. In Figure \ref{fig:lam_M1} we plot the distribution and moments of the non-zero eigenvalue for a simple fixed vector $\bm z^T = (0.25, 0.50, 0.75)$ and a range of vectors $\bm k$. This Figure helps to illustrate the fact that, unsurprisingly, changing the entries of the vector $\bm k$ has the effect of shifting the distribution towards the $z_i$ associated with the largest $k_i$.

\begin{thm}\label{thm:sig2_M1}
    Let $M_1 = \bm z \bm a^T$ be a random $N\times N$ rank-$1$ NNEM drawn according to Definition \ref{def:M1}. The non-negative integer moments ($m \geq 0$) of the non-zero squared singular value of $M_1$, given in Eq. \eqref{eq:lambda_svs_def} as $\sigma^2 = r^2 \varphi$ with $r^2 \equiv \bm z^T \bm z$ and $\varphi \equiv \bm a^T \bm a$, can be written as $\langle{(\sigma^2)}^m\rangle_{\bm a} = r^{2m} \expval{\varphi^m}_{\bm a}$, such that 
    \begin{equation}
        \expval{\varphi^m}_{\bm a} = \frac{1}{B(\bm k)} \frac{1}{\Gamma(\kappa + 2m)} \lim_{ w \to 0 } \, \frac{\diff^{m}}{\diff w^{m}} \left[ \prod_{i=1}^N \left( \sum_{j = 0}^m \frac{\Gamma(2j + k_i)}{\Gamma(j + 1)} w^{j} \right) \right] \ ,
        \label{eq:moms_sig2_GDRV}
    \end{equation}
    where $\kappa \equiv \sum_i k_i$ and $B(\bm k)$ is a normalisation constant defined in Eq. \eqref{eq:jpdf_GDRV}. Additionally, it follows that, for flat Dirichlet random variables ($k_i = 1$ for $i=1, 2, \ldots, N$), the integer moments of the squared non-zero singular value can be obtained through
    \begin{equation}
        \expval{ \varphi^m }_{\bm a} = \frac{\Gamma(N)}{\Gamma(N + 2m )} \sum\limits_{\ell =1}^{\min(N,m)} \frac{N!}{(N-\ell)! } \, B_{m,\ell}\Big( f(1), 2!f(2), \ldots, (m-\ell + 1)!f(m-\ell + 1)  \Big) 
        \label{eq:moms_sig2_FDRV}
    \end{equation}
    where $f(n) \equiv (2n)!/n!$ and $B_{m,\ell}(\cdot)$ are the Bell polynomials \cite{Wang09}.
\end{thm}

\begin{figure}[h]
  \begin{minipage}[h]{.45\linewidth}
    \centering
    \includegraphics[scale = 0.3]{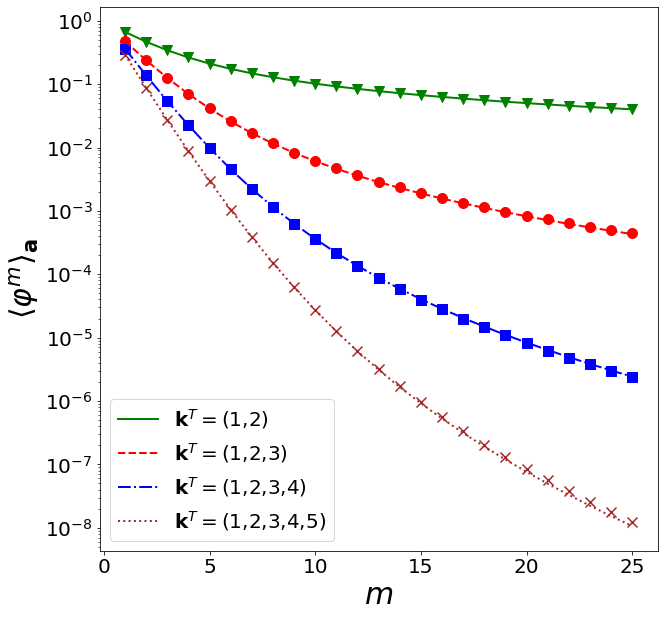}
  \end{minipage}\hspace{0.5cm}
  \begin{minipage}[h]{.45\linewidth}
    \centering
    \footnotesize 
    \begin{tabular}{|r|c|c|}
    \hline
    $m$ & Theory - Eq. \eqref{eq:moms_sig2_FDRV} & Simulation \\ \hline \hline

    $(N=3)$ 1 & $\frac{1}{2} = 5.000 \times 10^{-1}$ & $5.001 \times 10^{-1}$ \\ \hline
    2 & $\frac{4}{15} = 2.667 \times 10^{-1}$ & $2.668 \times 10^{-1}$ \\ \hline
    3 & $\frac{16}{105} = 1.524 \times 10^{-1}$ & $1.525 \times 10^{-1}$ \\ \hline
    4 & $\frac{7}{75} = 9.333 \times 10^{-2}$ & $9.341 \times 10^{-2}$\\ \hline \hline
    
    $(N=5)$ 1 & $\frac{1}{3} = 3.333 \times 10^{-1}$ & $3.333 \times 10^{-1}$ \\ \hline
    2 & $\frac{5}{42} = 1.190 \times 10^{-1}$ & $1.191 \times 10^{-1}$ \\ \hline
    3 & $\frac{29}{630} = 4.603 \times 10^{-2}$ & $4.605 \times 10^{-2}$ \\ \hline
    4 & $\frac{202}{10395} = 1.943 \times 10^{-2}$ & $1.945 \times 10^{-2}$\\ \hline \hline

    $(N=10)$ 1 & $\frac{2}{11}$ = $1.818\times 10^{-1}$ & ~ $1.818\times 10^{-1}$ ~ \\ \hline
    2 & $\frac{5}{143}$ = $3.497\times 10^{-2}$ & ~ $3.495\times 10^{-2}$ ~ \\ \hline
    3 & $\frac{36}{5005}$ = $7.193\times 10^{-3}$ & ~ $7.188\times 10^{-3}$ ~ \\ \hline
    4 & $\frac{3}{1870}$ = $1.604\times 10^{-3}$ & ~ $1.602\times 10^{-3}$ ~ \\ \hline 
    
    \end{tabular}
  \end{minipage}
  \caption{\small Non-negative integer moments, $m$, of the sum of the squares of Dirichlet random variables, $\varphi$, for a range of $N$ and $\bm k$. Left: Integer moments of $\varphi$ for $N=2,3,4$ and $5$ with $k_i = i$, theoretical lines are plotted according to Eq. \eqref{eq:moms_sig2_GDRV} and coloured markers are numerical averages. Right: Fractional values of expected moments of the sum of squares of flat Dirichlet random variables (generated from Eq. \eqref{eq:moms_sig2_FDRV}) and compared to numerical simulation for $N=3, 5$ and $10$. Numerical averages are calculated using the sum of the squares obtained from $10^6$ sets of DRVs drawn with the associated Dirichlet parameters. }
  \label{fig:svs_M1}
\end{figure}

\noindent
The results reported in Theorem \ref{thm:sig2_M1} are proved in Section \ref{sec:proofs_sig_M1} and are plotted alongside results from numerical simulations in Figure \ref{fig:svs_M1}. One should note that  computation of the distribution of the sum of squares of DRVs has proved too technically challenging. In spite of the apparent simplicity of formulating the problem, derivation of an exact expression is very cumbersome, and seems essentially intractable in practice. However, some results do exist in this area of study, see \cite{Royen_2007, Royen_2010}. 

\subsection{Results for rank-2 NNEMs}
\label{sec:Res_M2}

In order to state our results which refer to the product of the non-zero eigenvalues and their associated jpdf, we need to consider the joint statistics of the objects $\alpha, \beta, \gamma$ and $\zeta$, as defined in Eqs. \eqref{eq:T_def} and \eqref{eq:D_def}. Since the vectors $\bm a$ and $\bm b$ are independent, it follows that the only correlated variables are the pairs $(\alpha,\beta)$ and $(\gamma,\zeta)$. Moreover, up to exchanging $\bm h$ and $\bm k$, the joint statistics of $(\alpha,\beta)$ will be identical to those of $(\gamma, \zeta)$. To this end, we have calculated the joint distribution of $\alpha$ and $\beta$ and describe the results in the following proposition. This result can then be utilised to derive a range of important spectral quantities. 
\begin{prop} \label{prop:jpdf_alpha_beta}
    Let $\bm a$ be a vector of $N$ general Dirichlet random variables, associated with a set of Dirichlet parameters $k_i$, which form the entries of the vector $\bm k$. Let $\alpha = \bm a^T \bm x$ and $\beta = \bm a^T \bm y$, where $\bm x$ and $\bm y$ are two independent fixed non-negative vectors of size $N$. The jpdf of $\alpha$ and $\beta$ is given by 
    \begin{align}
        \overline{\mathcal{P}}_{\bm x, \bm y, \bm k}&(\alpha, \beta) = \nonumber \\ 
        &\begin{cases}
            \frac{1}{Y_{21}} \delta\left( \alpha - x_2 - \frac{X_{21}}{Y_{21}} (\beta - y_2) \right) \bigg( \Theta[\beta - y_1] - \Theta[\beta - y_2] \bigg) &[\kappa = 2]\\
            \frac{\Gamma(\kappa)(-1)^{\kappa-1}}{\Gamma(\kappa-2)} \bm \partial_{\bm v}^{\bm k} \Bigg[ \sum\limits_{i=1}^N \sum\limits_{j > i} \frac{Y_{ij}^{N-3}}{\prod\limits^N_{\ell = 1, \ell \neq j,k} \Big( \Delta_{j \ell} + \Delta_{ij} + \Delta_{\ell i} \Big) }  \bigg( \Theta[\beta - y_i]  - \Theta[\beta - y_j] \bigg)  \\
            \hspace{2.0cm} \times \left( \alpha - v_i - \frac{V_{ij}}{Y_{ij}}(\beta - y_i) \right)^{\kappa-3} \Theta\left[ \alpha - v_i - \frac{V_{ij}}{Y_{ij}}(\beta - y_i) \right] \Bigg]_{\bm v = \bm x} &[\kappa \geq 3] \ ,
        \end{cases} \label{eq:jpdf_alpha_beta_N} 
    \end{align}
    where $\kappa \equiv \sum_i k_i$, $\bm v = (v_1, v_2,\ldots ,v_N)^T$, and $\bm \partial_{\bm v}^{\bm k}$, defined according to Eq. \eqref{eq:def_partials}, is a set of $\kappa-N$ derivatives over the entries of $\bm v$ which is set equal to $\bm x$ after the derivatives have been taken. In this expression we have introduced the notations $V_{ij} \equiv v_i - v_j$, $Y_{ij} \equiv y_i - y_j$ and $\Delta_{ij} \equiv y_i v_j - y_j v_i$, which are all antisymmetric under the exchange of the indices $i$ and $j$. 
\end{prop}

\noindent
The Proposition is verified in Section \ref{sec:proof_jpdf_alpha_beta}. With this result now in place, we are able to present our results for the non-zero eigenvalues of the matrix $M_2$.

\begin{thm} \label{thm:lam_M2}
    Let $M_2$ be a random $N\times N$ rank-$2$ NNEM drawn according to Definition \ref{def:M2}. The joint distribution of the sum and product of the eigenvalues of $M_2$, defined in Eqs. \eqref{eq:T_def} and \eqref{eq:D_def} respectively, is given by 
    \begin{align}
        \mathcal{P}_{\bm x, \bm y, \bm h, \bm k}(T, D) = \int_0^\infty \diff \alpha \int_0^\infty \frac{\diff \beta}{\beta} \, &\overline{\mathcal{P}}_{\bm x, \bm y, \bm h}(\alpha, \beta) \overline{\mathcal{P}}_{\bm x, \bm y, \bm k}\left( \frac{\alpha(T- \alpha) - D}{\beta}, T - \alpha \right)  \nonumber \\
        &\times \Theta\left[ \frac{\alpha(T- \alpha) - D}{\beta}\right] \Theta\left[  T - \alpha \right] \ , \label{eq:jpdf_T_D}
    \end{align}
    where $\overline{\mathcal{P}}_{\bm x, \bm y, \bm k}(\alpha, \beta)$ is given in Eq. \eqref{eq:jpdf_alpha_beta_N}. Further to this, the jpdf of the two non-zero eigenvalues, $\lambda_1$ and $\lambda_2$, is given by
    \begin{equation}
        \mathcal{P}_{\bm x, \bm y, \bm h, \bm k}^{(\lambda_1, \lambda_2)}(\lambda_1, \lambda_2) = \mathcal{P}_{\bm x, \bm y, \bm h, \bm k}\Big(\lambda_1 + \lambda_2, \lambda_1 \lambda_2 \Big) \Theta\Big[\lambda_1 + \lambda_2 \Big] \ .
    \end{equation}
    The marginal distribution of the sum of the non-zero eigenvalues of $M_2$, denoted by $T$, is given at finite $N \geq 2$ by
    \begin{align}
        \mathcal{P}_{\bm x, \bm y, \bm h, \bm k}&(T) \equiv \bigg\langle \delta\left( T - \bm a^T \bm x - \bm b^T \bm y\right) \bigg\rangle_{\bm a, \bm b} = \frac{\Gamma(\kappa) \, \Gamma(\eta) }{ \Gamma(\kappa + \eta - 2) } \nonumber  \\
        &  \times \bm \partial_{\bm x_1}^{\bm h} \bm \partial_{\bm x_2}^{\bm k} \Bigg[ \sum_{j,m = 1}^N 
         \frac{( x_{1j} + x_{2m} - T )^{\kappa + \eta - 3}}{  \prod\limits_{\ell \neq j} (x_{1j} - x_{1 \ell}) \prod\limits_{n \neq m} (x_{2m} - x_{2n})  } \Theta[ x_{1j} + x_{2m} - T] \Bigg]_{ \underset{\bm x_2 = \bm y}{\bm x_1 = \bm x }}\ ,  \label{eq:pdf_T_M2}
    \end{align}
    where $\bm \partial_{\bm x}^{\bm k}$ is defined in Eq. \eqref{eq:def_partials} and $x_{1j}$ and $x_{2m}$ are the $j$th and $m$th entries of $\bm x_1$ and $\bm x_2$ respectively. Furthermore, the associated non-negative integer moments ($m\geq 0$) of $T$, defined in Eq. \eqref{eq:T_def}, can be obtained from the above distribution as
    \begin{align}
        \expval{T^m}_{\bm a, \bm b} \equiv& \int_0^\infty \diff T \, T^m \, \mathcal{P}_{\bm x, \bm y, \bm h, \bm k}(T) \nonumber\\
        =& \frac{\Gamma(\eta) \, \Gamma(\kappa)  \, \Gamma(m+1)}{ \Gamma(\eta + \kappa + m - 1) } \bm \partial_{\bm x_1}^{\bm h} \bm \partial_{\bm x_2}^{\bm k} \left[ \sum_{j,m = 1}^N \frac{( x_{1j} + x_{2m} )^{\kappa + \eta - 2}}{  \prod\limits_{\ell \neq j} (x_{1j} - x_{1 \ell}) \prod\limits_{n \neq m} (x_{2m} - x_{2n})  }  \right]_{  
        \underset{\bm x_2 = \bm y}{\bm x_1 = \bm x} }\ .
        \label{eq:moms_T_M2}
    \end{align}
    Additionally, the equation for the non-negative integer moments of the product of the eigenvalues, denoted by $D$ and defined in Eq. \eqref{eq:D_def}, is given by
    \begin{equation}
        \expval{D^m}_{\bm a, \bm b} = \frac{1}{B(\bm h) B(\bm k)} \sum_{n = 0}^m \frac{m! \, n! \, (m-n)! \, (-1)^{m-n}}{\Gamma(m + \eta) \Gamma(m + \kappa)} \, P_{\bm x, \bm y, \bm h}(n, m - n) \, P_{\bm y, \bm x, \bm k}(n, m - n) \ , 
        \label{eq:moms_D_GDRV}
    \end{equation}
    with
    \begin{equation}
        P_{\bm x, \bm y, \bm h}(m_1, m_2) =
        \frac{1}{m_1!} \lim_{w \to 0} \frac{\diff^{m_1}}{\diff w^{m_1}} \left[ \sum_{n_1=0}^{m_1} \frac{(wx_1)^{n_1}}{n_1!} \cdots \sum_{n_N=0}^{m_1} \frac{(wx_N)^{n_N}}{n_N!} \, Q_{\bm y, \bm h}(m_2, \bm n) \right]
    \end{equation}
    and 
    \begin{equation}
        Q_{\bm y, \bm h}(m_2, \bm n) = \frac{1}{m_2!} \lim_{w \to 0} \frac{\diff^{m_2}}{\diff w^{m_2}} \left[ \prod_{i=1}^N \left( \sum_{j=0}^{m_2} \frac{(wy_i)^j \, \Gamma(n_i + h_i + j) }{j!} \right) \right] \ .
    \end{equation}
\end{thm}

\begin{figure}[h]
    \centering
    \includegraphics[scale=0.3]{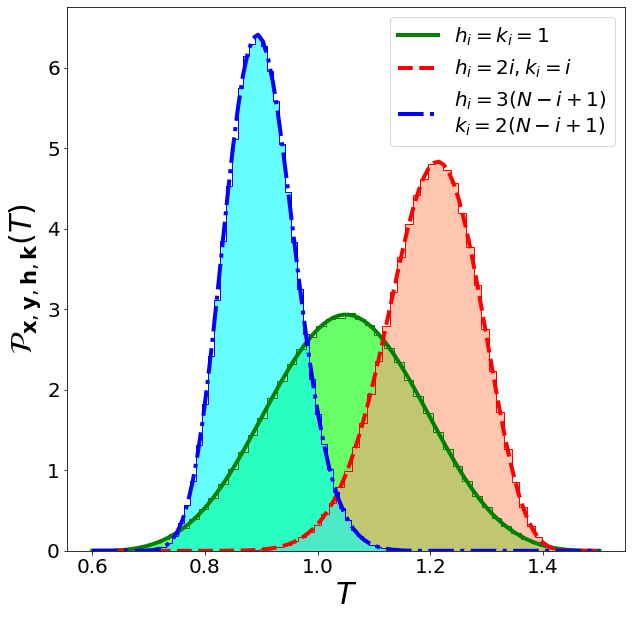}
    \hspace{1cm}
    \includegraphics[scale=0.3]{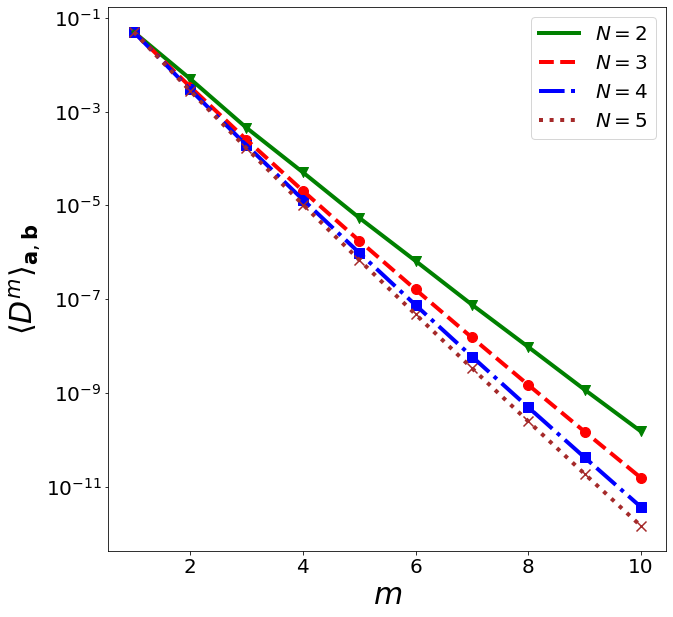}
    \caption{\small Statistics of the interplay between the non-zero eigenvalues of the rank-$2$ NNEM $M_2$, as defined in Eq. \eqref{eq:M2_def}. Left: Comparison between theoretical distributions of the sum of the non-zero eigenvalues (coloured lines plotted using Eq. \eqref{eq:pdf_T_M2}) and histograms of numerically generated data, for $N=3$ and a range of $\bm h$ and $\bm k$. Right: Non-negative integer moments of the product of the non-zero eigenvalues for $N = 2,3,4$ and $5$, using $h_i = i$ and $k_i = (N+1)-i$, coloured lines are theoretical predictions according to Eq. \eqref{eq:moms_D_GDRV} and coloured markers are numerical averages. In these plots the fixed vectors, $\bm x$ and $\bm y$, have linearly spaced entries on $[0.1,0.5]$ and $[0.5,1]$ respectively. Data from numerical simulation consists of $10^6$ random samples of the eigenvalues.} 
    \label{fig:M2_lams}
\end{figure}

\noindent
For the proof of these results, one should consult Section \ref{sec:proofs_lam_M2}. In Figure \ref{fig:M2_lams} we plot the results of Theorem \ref{thm:lam_M2} for the sum and product of the non-zero eigenvalues of $M_2$ and compare them to numerical simulations for a range of $N$ and choices of $\bm h$ and $\bm k$.

\begin{thm}\label{thm:sig2_M2}
    Let $M_2$ be a random $N \times N$ rank-$2$ NNEM drawn according to Definition \ref{def:M2}. The non-negative integer moments ($m\geq 0$) of the sum of the squared singular values of $M_2$, defined in Eq. \eqref{eq:tau_def}, are given by
    \begin{align}
        \expval{\tau^m}_{\bm a, \bm b} = \frac{m!}{B(\bm h) B(\bm k)} \Bigg[ \sum_{i=0}^m \, \sum_{j=0}^m & \frac{r_{xx}^{i} \, r_{yy}^{j} \, (2 r_{xy})^{m-(i+j)} }{\Gamma(m + i - j + \eta) \, \Gamma(m + j - i + \kappa)} \nonumber \\
        & \times \chi_{\bm h, \bm k} \Big(i, j, m - (i+j) \Big) \Theta\Big[ m-(i+j) \Big] \Bigg] \label{eq:moms_tau_M2} \ , 
    \end{align}
    such that $r_{xx} \equiv \bm x^T \bm x$, $r_{yy} \equiv \bm y^T \bm y$ and $r_{xy} \equiv \bm x^T \bm y$, additionally
    \begin{equation}
        \chi_{\bm h, \bm k}(m_1, m_2, m_3) = \frac{1}{m_3!} \lim_{w \to 0} \frac{\diff^{m_3}}{\diff w^{m_3}} \left[ \sum_{r_1 = 0}^{m_3} \frac{w^{r_1}}{r_1!} \sum_{r_2 = 0}^{m_3} \frac{w^{r_2}}{r_2!}\cdots \sum_{r_N = 0}^{m_3} \frac{w^{r_N}}{r_N!} S_{\bm h}(m_1, \bm r) S_{\bm k}(m_2, \bm r)  \right]
        \label{eq:chi_3args}
    \end{equation}
    where $\bm r = (r_1, r_2, \ldots , r_N)^T$ is a vector of integers and
    \begin{equation}
        S_{\bm h}(m, \bm r) = \frac{1}{m!} \lim_{w \to 0} \frac{\diff^m}{\diff w^m} \left[ \prod_{i=1}^N \left( \sum_{j=0}^m \frac{\Gamma(2j + r_i + h_i)}{j!} w^{j} \right) \right] \ .
    \end{equation}
    Furthermore, the corresponding moments of the product of the squared singular values, given in Eq. \eqref{eq:Delta_def} as $\Delta = \Delta_{xy} \Delta_{ab}$, where $\Delta_{a b} \equiv (\bm a^T \bm a) (\bm b^T \bm b) - (\bm a^T \bm b)^2$ and $\Delta_{x y} \equiv (\bm x^T \bm x) (\bm y^T \bm y) - (\bm x^T \bm y)^2$, can be written as $\expval{\Delta^m}_{\bm a, \bm b} = \Delta_{x y}^m \expval{\Delta_{a b}^m}_{\bm a, \bm b}$, such that
    \begin{equation}
        \expval{\Delta_{ab}^m}_{\bm a, \bm b} = \frac{m!}{B(\bm h) B(\bm k)} \sum_{n=0}^m \frac{(-1)^{m-n} \, n! \, (2(m-n))!}{(m-n)! \, \Gamma(2m + \eta)  \, \Gamma(2m + \kappa)} \, \chi_{\bm h, \bm k}\Big(n, n, 2 (m-n) \Big) \ ,
        \label{eq:moms_Delta_ab}
    \end{equation}
    with $\chi_{\bm h, \bm k}(\cdot)$ defined in Eq. \eqref{eq:chi_3args}. 
\end{thm}

\begin{figure}[h]
    \centering
    \includegraphics[scale = 0.3]{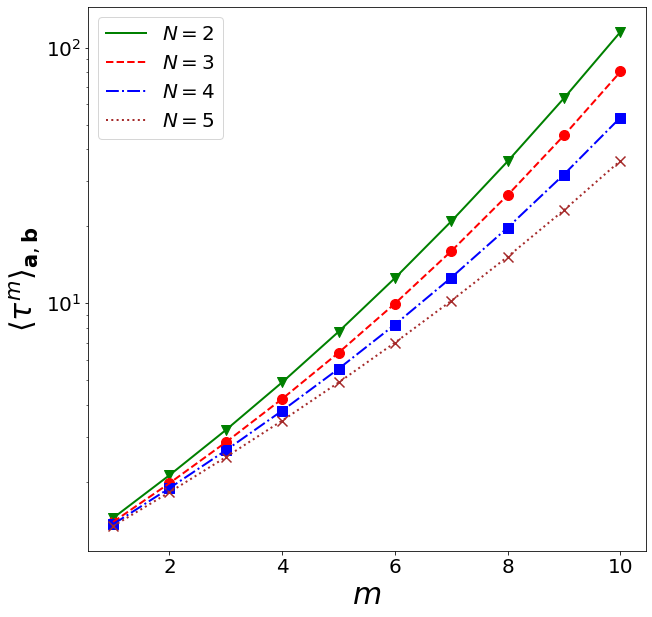}
    \hspace{1cm}
    \includegraphics[scale=0.3]{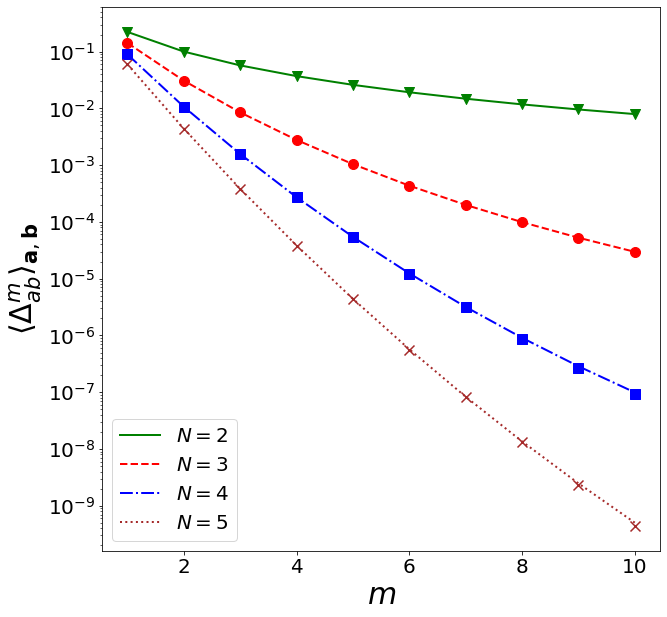}
    \caption{\small Non-negative integer moments, $m$, of the interplay between the non-zero squared singular values of the rank-2 NNEM $M_2$, defined in Eq. \eqref{eq:M2_def}. In both plots we choose the entries of $\bm h$ and $\bm k$ such that $h_i = i$ and $k_i = N+1-i$ for $i = 1,2,\ldots, N$. Left: Integer moments of the sum of the squared singular values using linearly spaced entries of $\bm x \in [0, 0.5]$ and $\bm y \in [0.5, 1]$, plotted according to Eq. \eqref{eq:moms_tau_M2} (coloured lines). Right: Integer moments of $\Delta_{ab}$ which is central to determining the moments of the product of the squared singular values, plotted using Eq. \eqref{eq:moms_Delta_ab} (coloured lines). Numerical averages (coloured markers) are obtained using a sample of $10^6$ randomly generated sets of Dirichlet random variables.}
    \label{fig:svs_M2}
\end{figure}

\noindent
Detailed proofs of these results can be found in Section \ref{sec:proofs_sig_M2}. In Figure \ref{fig:svs_M2}, we plot some examples of the moments of the sum and product of the non-zero squared singular values of $M_2$ and compare to numerical simulation for a range of $N$ and choices of $\bm h$ and $\bm k$.

\subsection{Results for rank-$R$ NNEMs}
\label{sec:Res_MR}

In this section, we provide some results for the statistics of the non-zero eigenvalues and squared singular values of the rank-$R$ NNEM $M_R$, starting with the eigenvalues. One should note that all results in this section can be applied to full-rank matrices by setting $R=N$. \\

\begin{thm}\label{thm:lam_MR}
    Let $M_R$ be a random rank-$R$ NNEM of size $N \times N$ drawn according to Definition \ref{def:MR}. In this setting, the distribution of the sum of the non-zero eigenvalues of $M_R$, denoted as $T_R$ and defined in Eq. \eqref{eq:def_T_D_MR}, is given by
    \begin{align}
        \mathcal{P}_{Z, K}(T_R) \equiv& \expval{\delta\left(T_R - \sum_{i=1}^R \bm a_i^T \bm z_i \right)}_A = \frac{\left(\prod\limits_{i=1}^R \Gamma(\kappa_i)\right) }{\Gamma(\widehat{\kappa}_R - R)} \bm \partial_{X}^K \Bigg[ \sum_{j_1=1}^N \omega_{1j_1} \sum_{j_2=1}^N \omega_{2j_2} \cdots \sum_{j_R=1}^N \omega_{Rj_R} \nonumber\\
        &\times \left(  \sum_{i=1}^R X_{ij_i} - T_R \right)^{\widehat{\kappa}_R - R -1} \Theta\left[ \sum_{i=1}^R X_{ij_i} - T_R \right] \Bigg]_{X=Z} \ , \label{eq:pdf_T_MR}
    \end{align}
    such that
    \begin{equation}
        \widehat{\kappa}_R = \sum_{i=1}^R \kappa_i \ , \hspace{0.5cm}  \omega_{ij_i} = \left(\prod\limits_{\ell \neq j_i}^N (X_{ij_i} - X_{i \ell})\right)^{-1} \ , \hspace{0.5cm} \bm \partial_X^{K} = \prod_{i=1}^R \prod_{j=1}^N \frac{1}{\Gamma(K_{ij})} \frac{\partial^{K_{ij} - 1}}{\partial X_{ij}^{K_{ij} - 1}}\ ,
    \end{equation}
    where $X$ is defined analogously to $Z$ and $K$, introduced in Eq. \eqref{eq:ZKA_def}. Additionally, one can utilise the above pdf to write down an expression for the non-negative integer moments ($m\geq 0$) of $T_R$, which reads
    \begingroup
    \allowdisplaybreaks
    \begin{align}
        \expval{T_R^m}_A &\equiv \int_0^\infty \diff T_R \, T_R^m \, \mathcal{P}_{Z, K}(T_R) \nonumber \\
        =& \frac{\left(\prod\limits_{i=1}^R \Gamma(\kappa_i)\right) \Gamma(m+1) }{\Gamma(\widehat{\kappa}_R + m + 1 - R)} \bm \partial_{X}^K \left[ \sum_{j_1=1}^N \omega_{1j_1} \cdots \sum_{j_R=1}^N \omega_{Rj_R} \left( \sum_{i=1}^R X_{ij_i} \right)^{\widehat{\kappa}_R + m - R}  \right]_{X=Z}   .
        \label{eq:moms_T_MR}
    \end{align}
    \endgroup
    Finally, the first moment of the product of the eigenvalues, also defined in Eq. \eqref{eq:def_T_D_MR}, is given by
    \begin{equation}
        \expval{D_R}_A =  \left( \prod_{i=1}^R \frac{1}{\kappa_i} \right) \det\bigg( KZ^T \bigg) \ ,
        \label{eq:mom1_DR}
    \end{equation}
    where the $(i,j)$th entry of the matrix $KZ^T$ is given by $(KZ^T)_{ij} = \bm k_i^T \bm z_j$ for $i,j = 1,2, \ldots, R$.
\end{thm}

\noindent
For the proof of these results see Section \ref{sec:proofs_lam_MR}.

\begin{figure}[h]
  \begin{minipage}[h]{.45\linewidth}
    \centering
    \includegraphics[scale = 0.3]{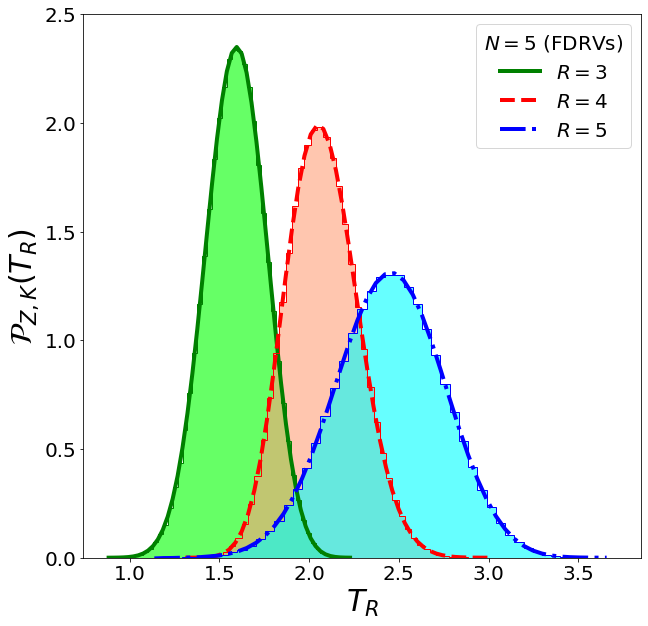}
  \end{minipage}\hspace{0.5cm}
  \begin{minipage}[h]{.45\linewidth}
    \centering
    \footnotesize 
    \begin{tabular}{|r|c|c|}
    \hline
    $\expval{D_R}_A \quad R$ & Theory - Eq. \eqref{eq:mom1_DR} & Simulation \\ \hline \hline

    3 & $-1.535 \times 10^{-3}$ & $-1.529 \times 10^{-3}$ \\ \hline
    4 & $ 4.702 \times 10^{-4}$ & $ 4.707 \times 10^{-4}$ \\ \hline
    5 & $-1.210 \times 10^{-6}$ & $-1.215 \times 10^{-6}$ \\ \hline \hline
    
    $\expval{\tau_R}_A \quad R$ & Theory - Eq. \eqref{eq:mom1_tauR} & Simulation \\ \hline \hline
    3 & $3.187$ & $3.187$ \\ \hline
    4 & $4.046$ & $4.045$ \\ \hline
    5 & $8.936$ & $8.935$\\ \hline \hline

    $\expval{\Sigma_A}_A \quad R$ & Theory - Eq. \eqref{eq:mom1_DeltaR} & Simulation \\ \hline \hline
    3 & $3.329 \times 10^{-3}$ & $3.329 \times 10^{-3}$ \\ \hline
    4 & $2.320 \times 10^{-4}$ & $2.324 \times 10^{-4}$ \\ \hline 
    
    \end{tabular}
  \end{minipage}
  \caption{\small Statistics of the sum and product of the non-zero eigenvalues and singular values of $M_R$. In each dataset, entries of $Z$ are drawn from a flat distribution on $[0,1]$ and $N=5$. Left: Distribution of the sum of the eigenvalues (coloured lines plotted according to Eq. \eqref{eq:pdf_lambda_GDRV}) compared to numerical simulation (histograms)  for: $R=3$ (green), $R=4$ (red) and $R=5$ (blue) - in all plots the DRVs are flat. Right: First moment of: $D_R$ (top table), $\tau_R$ (middle table) and $\Sigma_A$ (bottom table) for a selection of different values of $R$. In each dataset of the table the fixed entries of $K$ are drawn from flat integer distributions between $[1,5]$. Exact simulation parameters are available upon request. Each data set is generated using $10^6$ random samples of $R$ independent vectors of DRVs. }
  \label{fig:MR}
\end{figure}

\begin{thm} \label{thm:sig_MR}
    Let $M_R$ be a random rank-$R$ NNEM of size $N \times N$ drawn according to Definition \ref{def:MR}. The first moment of the sum of the squared singular values of $M_R$, defined in Eq. \eqref{eq:def_tau_MR}, is given by
    \begin{equation}
        \expval{\tau_R}_A = \sum_{i,j=1}^N \expval{\bm a_i^T \bm a_j}_{\bm a_i , \bm a_j } \bm z_j^T \bm z_i \hspace{1cm} \text{where} \hspace{1cm} \expval{\bm a_i^T \bm a_j}_{\bm a_i, \bm a_j} = \begin{cases}
            \frac{\bm k_i^T \bm k_i + \kappa_i}{ \kappa_i (\kappa_i + 1)} \hspace{1cm}&[i = j] \\
            \frac{\bm k_i^T \bm k_j}{ \kappa_i \kappa_j} \hspace{1cm}&[i \neq j]
        \end{cases} \ .
        \label{eq:mom1_tauR}
    \end{equation}
    Furthermore, using Eq. \eqref{eq:def_Delta_MR}, the first moment of  the product of the non-zero squared singular values of $M_R$ can be written as
    \begin{equation}
        \expval{\Delta_R}_{A} = \det(\Sigma_Z) \expval{\det(\Sigma_A)}_{A} \ ,
        \label{eq:mom1_DeltaR}
    \end{equation}
    with $\Sigma_A \equiv AA^T$ and $\Sigma_Z \equiv ZZ^T$. The average determinant of this matrix is hard to present in a closed form, however it can be obtained using combinations of the object
    \begin{equation}
        F_K\left( \{ \ell_1, \ell_2, \ldots, \ell_s \} \right) \equiv \expval{(\bm a_{\ell_1}^T \bm a_{\ell_2}) (\bm a_{\ell_2}^T \bm a_{\ell_3}) \cdots (\bm a_{\ell_s}^T \bm a_{\ell_1})}_{\bm a_{\ell_1},\ldots ,\bm a_{\ell_s}} = \Tr\left( \prod_{i=1}^s \Psi_{\ell_i} \right) \ , 
    \end{equation}
    where $1 \leq s \leq R$ and $\Psi_{\nu} \equiv \expval{\bm a_{\nu} \bm a_{\nu}^T}_{\bm a_{\nu}}$, for $\nu = 1,2,\ldots,R$, with entries given by
    \begin{equation}
        \big( \Psi_{\nu} \big)_{ij} = \frac{1}{\kappa_{\nu}(\kappa_{\nu} + 1)}\begin{cases}
            K_{\nu i} (K_{\nu i} + 1) \hspace{2cm}& [i=j] \\
            K_{\nu i} K_{\nu j} \hspace{2cm}& [i\neq j]
        \end{cases} \ .
    \end{equation}    
    For an example of this process, one can consult Section \ref{sec:proofs_sig_MR}. 
\end{thm}
\noindent
For a proof of these results we refer the reader to Section \ref{sec:proofs_sig_MR}. Furthermore, in Figure \ref{fig:MR}, we compare our theoretical results for the eigenvalues and singular values of $M_R$ to results of direct numerical simulation.

\subsection{Discussion \& outlook}
\label{sec:Discussion}

Matrices with non-negative entries and prescribed row sums have far-reaching applications in a wide range of fields including complex networks and economics. However, in spite of their deep-rooted presence in many research areas, relatively little is known analytically about their spectra and Extreme Values thereof. The positivity constraint breaks any hope to preserve rotational invariance and rules out the use of Gaussian integral representations, thus dramatically reducing the analytical arsenal available to deal with them. The primary aim of this work was to provide some analytic insight into the spectra and statistical properties of such matrices through the use of random low-rank approximations.  

We have been able to accurately derive a range of results pertaining to the sum and product of the non-zero eigenvalues and singular values of low-rank NNEMs, particularly focusing on rank-$1$ and rank-$2$ but also deriving results for a general rank-$R$. The random matrix models studied in this work provide rare and useful examples of systems where the full Extreme Value Statistics can be computed analytically for finite matrix size. 

The work presented here leads to a few questions worthy of further research. In particular, we have been able to present low-rank matrix approximations (potentially preserving macroscopic properties of a full-rank matrix) for which the full Extreme Value Statistics can be derived. This leads one to naturally question whether there are more low-rank approximations that can also be fully analysed? Fruitful extensions could include the preservation of more macroscopic quantities (in addition to prescribed row sums) such as the: largest eigenvalues, spectral gap and sums of the elements in each column. One could also consider methods to optimise an approximation, such that it minimises the difference between true macroscopic quantities and the ones seen in the approximation. Furthermore, one could also consider imposing further constraints on the rank-$R$ model studied in this work by fixing the row sum as the sum of $R$ random variables. Specifically, in our current model the sum of the elements in row $i$ is given by $\sum_{\ell=1}^R (\bm z_\ell)_i$, but this could be generalised by constraining each vector such that $\sum_{\ell=1}^R (\bm z_\ell)_i = z_i$. 

\section{Proofs of results for rank-1 NNEMs}
\label{sec:rank1_proofs}

In this section, we give proofs of the results related to the single non-zero eigenvalue and singular value of the rank-$1$ NNEM $M_1$. We start in Section \ref{sec:proofs_lam_M1} by proving the equations for the non-zero eigenvalue's distribution and moments and then consider the moments of the singular value in Section \ref{sec:proofs_sig_M1}.

\subsection{Non-zero eigenvalue of rank-$1$ NNEMs}
\label{sec:proofs_lam_M1}

We now provide proofs of the results given in Theorem \ref{thm:lam_M1}, specifically Eqs. \eqref{eq:pdf_lambda_GDRV} \& \eqref{eq:moms_lambda_GDRV}, which describe the distribution and moments of the non-zero eigenvalue of $M_1$.
\begin{proof}
    To prove Eq. \eqref{eq:pdf_lambda_GDRV} we begin by employing the definition of the general Dirichlet random vector from Eq. \eqref{eq:jpdf_GDRV}, so as to write that
    \begin{equation}
        \mathcal{P}_{\bm z, \bm k}(\lambda, t) \equiv \frac{1}{B(\bm k)} \left[ \prod_{i=1}^{N}  \int_0^\infty \diff a_i \, a_i^{k_i - 1} \right] \delta \left( \sum_{i=1}^N a_i - t \right) \delta \left( \lambda - \sum_{i=1}^N a_i z_i \right) \ ,
    \end{equation}
    where we have introduced the auxiliary variable $t$ for the sum of the DRVs, which we later set to 1. Setting $t=1$ gives the distribution of $\lambda$, denoted as $\mathcal{P}_{\bm z, \bm k}(\lambda) \equiv \mathcal{P}_{\bm z, \bm k}(\lambda,1)$. One can then take a double Laplace transform w.r.t. $\lambda$ and $t$ and see that
    \begin{align}
        \widetilde{\mathcal{P}}^{(2)}_{\bm z, \bm k}(\mu, s) \equiv & \int_0^\infty \diff \lambda \, e^{-\lambda \mu} \int_0^\infty \diff t \, e^{-st} \, \mathcal{P}_{\bm z, \bm k}(\lambda, t) \nonumber\\
        =& \frac{1}{B(\bm k)} \left[ \prod_{i=1}^{N} \int_0^\infty \diff a_i \, a_i^{k_i - 1} \, e^{-(s + \mu z_i) a_i} \right] = \Gamma(\kappa) \prod_{i=1}^N \frac{1}{(s+\mu z_i)^{k_i}} \ ,
    \end{align}
    where we have employed the definition of $B(\bm k)$ from Eq. \eqref{eq:jpdf_GDRV} and recall that $\kappa = \sum_i k_i$. Note that here the superscript $(2)$ denotes that we have 2 variables in Laplace space. In general, throughout this manuscript, a superscript of this form will be utilised to denote the number of variables in Laplace space and later a colon will be used to separate variables in Laplace space from those in usual space. One now notices that 
    \begin{equation}
        \frac{1}{(s+\mu z_i)^{k_i}} = \frac{(-1)^{k_i-1}}{\Gamma(k_i) \, \mu^{k_i-1}} \frac{\partial^{k_i-1}}{\partial x_i^{k_i - 1}} \left( \frac{1}{s + \mu x_i} \right) \Bigg|_{x_i = z_i} \ ,  
    \end{equation}
    which allows us to write that
    \begin{equation}
         \widetilde{\mathcal{P}}^{(2)}_{\bm z, \bm k}(\mu, s) =  \frac{\Gamma(\kappa)(-1)^{\kappa-N}}{\mu^{\kappa-N}} \, \bm \partial^{\bm k}_{\bm x} \, \left[ \prod_{i=1}^N \frac{1}{s + \mu x_i} \right]_{\bm x = \bm z} \hspace{0.25cm} \text{where} \hspace{0.5cm} \partial^{\bm k}_{\bm x} = \prod_{i=1}^N \frac{1}{\Gamma(k_i)} \frac{\partial^{k_i-1}}{\partial x_i^{k_i-1}} \ .
    \end{equation}
    If we now utilise the partial fraction decomposition
    \begin{equation}
        \prod_{i=1}^N \frac{1}{s + \mu x_i} = \sum_{j=1}^N \frac{x_j^{N-1}}{\prod \limits_{\ell \neq j} (x_j - x_\ell)} \, \frac{1}{s^{N-1} (s + \mu x_j)} \ ,
    \end{equation}
    we arrive at 
    \begin{equation}
        \widetilde{\mathcal{P}}^{(2)}_{\bm z, \bm k}(\mu, s) = \Gamma(\kappa)(-1)^{\kappa-N} \, \bm \partial^{\bm k}_{\bm x} \, \left[ \sum_{j=1}^N \frac{x_j^{N-1}}{\prod \limits_{\ell \neq j} (x_j - x_\ell)} \, \frac{1}{s^{N-1} \mu^{\kappa - N} (s + \mu x_j)} \right]_{\bm x = \bm z} \ .
        \label{eq:pdf_lambda_sums}
    \end{equation}
    Thus one must now perform a double inverse Laplace transform (ILT), denoted by $\mathcal{L}^{-1}$, on an object of the form
    \begin{equation}
        \widetilde{F}^{(2)}(\mu,s;P,Q,x) \equiv \frac{1}{s^P \mu^Q (s + \mu x)} 
    \end{equation}
    for non-negative integers $P$ and $Q$ (later we set $P=N-1$ and $Q=\kappa - N$). We begin by performing the Laplace inversion w.r.t. $s$ and then set $t=1$, finding that
    \begin{equation}
        \widetilde{F}^{(1)}(\mu;P,Q,x) \equiv \mathcal{L}^{-1}_{s \to t}\left[\widetilde{F}^{(2)}(\mu,s;P,Q,x) \right] \Big|_{t=1} = \frac{(-1)^P}{ x^{P}} \frac{e^{-x \mu}}{\mu^{P+Q}} \left( 1 - \frac{\Gamma\left(P , - x \mu \right)}{\Gamma(P)}\right) \ ,
        \label{eq:1LT_proof_lambda}
    \end{equation}
    which is expressed in terms of the incomplete $\Gamma$-function
    \begin{equation}
        \Gamma\left( N , a \right) \equiv \Gamma\left(N\right) \, e^{-a} \, \sum_{k=0}^{N-1} \frac{a^k}{k!} =  \int_{a}^{\infty} \diff u \, u^{N-1} \, e^{-u} \ .
        \label{eq:incmpl_Gamma}
    \end{equation}
    We start by first tackling the term arising from the $1$ in the large bracket of Eq. \eqref{eq:1LT_proof_lambda},
    \begin{equation}
        \widetilde{F}_1^{(1)}(\mu; P, Q, x) \equiv \frac{(-1)^P}{ x^{P}} \frac{e^{-x \mu}}{\mu^{P+Q}} \ ,
    \end{equation}
    which can be easily inverted from Laplace space to see that
    \begin{equation}
        F_1(\lambda;P,Q,x) \equiv \mathcal{L}^{-1}_{\mu \to \lambda}\left[\widetilde{F}_1^{(1)}(\mu; P,Q,x) \right] = \frac{(-1)^P}{\Gamma(P+Q)} \frac{(\lambda - x)^{P+Q-1}}{x^P} \Theta[\lambda - x] \ . 
    \end{equation}    
    On the other hand, considering the term arising from the ratio of the $\Gamma$-functions in Eq. \eqref{eq:1LT_proof_lambda} and using the series representation of the incomplete $\Gamma$-function from Eq. \eqref{eq:incmpl_Gamma}, we can write that
    \begin{equation}
        \widetilde{F}_2^{(1)}(\mu;P,Q,x) \equiv \frac{(-1)^P}{ x^{P}} \frac{e^{-x \mu}}{\mu^{P+Q}}  \frac{\Gamma\left(P , - x \mu \right)}{\Gamma(P)} = (-1)^P \sum_{n=0}^{P-1} \frac{(-1)^n}{n!} \frac{x^{n-P}}{\mu^{P+Q-n}} \ .
    \end{equation}
    One may now sum the contributions from each of the terms arising from the different $x_j$s, exactly as described in Eq. \eqref{eq:pdf_lambda_sums}, to see that
    \begin{align}
        \bm \partial_{\bm x}^{\bm k} \Bigg[ \sum_{j=1}^N \frac{x_j^{N-1}}{\prod\limits_{\ell \neq j}(x_j - x_{\ell})}& \widetilde{F}_2^{(1)}(\mu;N-1,\kappa-N,x_j) \Bigg]  \nonumber  \\
        &=(-1)^{N-1} \sum_{n=0}^{N-2} \frac{(-1)^n}{n!} \frac{1}{\mu^{\kappa-1-n}} \bm \partial_{\bm x}^{\bm k} \left[ \sum_{j=1}^N \frac{x_j^{n}}{\prod\limits_{\ell \neq j}(x_j - x_{\ell})}  \right] \ .\label{eq:contrib_F2}
    \end{align}
    Now we recall the identity \cite{AER24}
    \begin{equation}
        S_m(\bm z) = \sum_{k=1}^N \frac{z_k^m}{\prod\limits_{\ell \neq k}(z_k - z_\ell)} = \begin{cases}
            0 & m = 0,1, \ldots , N-2 \\
            1 & m = N - 1 \\
            \sum\limits_{i_1,\ldots,i_N = 0}^{m-(N-1)} \left(\prod\limits_{j=1}^N z_j^{i_j} \right) \delta_{\sum_j i_j, m-(N-1) }\hspace{1cm} & m = N, N + 1, \ldots 
        \end{cases}
        \label{eq:S_m}
    \end{equation} 
    and furthermore one can show that  
    \begin{equation}
        \bm \partial_{\bm x}^{\bm k} \, S_m (\bm x) = 0 \hspace{2cm} \text{for } m = 0, 1, \ldots , \kappa - 2 \ . 
        \label{eq:deriv_Sm}
    \end{equation}
    Now returning to Eq. \eqref{eq:contrib_F2}, we notice that, as the exponents of $x_j$ range from $[0, N-2]$, the corresponding sums vanish according to Eq. \eqref{eq:S_m}. This implies that $F_1(\lambda;P,Q,x_j)$ is the only term non-trivially contributing to the pdf. Using $P=N-1$ and $Q=\kappa-N$ we can write that
    \begin{equation}
        \mathcal{P}_{\bm z, \bm k}(\lambda) = \frac{\Gamma(\kappa)(-1)^{\kappa-1}}{\Gamma(\kappa - 1)} \, \bm \partial_{\bm x}^{\bm k} \, \left[ \sum\limits_{j=1}^{N} \frac{\left( \lambda - x_j \right)^{\kappa-2}}{\prod\limits_{l\neq j}^N (x_j - x_l)}  \Theta\left[ \lambda - x_j \right] \right]_{\bm x = \bm z} \ .
        \label{eq:pdf_lambda_form1}
    \end{equation}
    At this point it is useful to note that one can fairly easily ascertain the range of the non-zero support of this pdf. Firstly, one can see that if we choose a value of $\lambda < \min(\bm z)$ then all the $\Theta$ functions tend to zero. Secondly, if one chooses a value of $\lambda > \max(\bm z)$ then all the $\Theta$ functions become 1. Utilising the binomial theorem, one can see that the greatest exponent of $x_j$ in the summation over $j$ will be $\kappa-2$ and so, by Eq. \eqref{eq:deriv_Sm}, this will also go to zero. Hence the only non-zero region of the pdf is when $\lambda \in [\min(\bm z), \max(\bm z)]$. Furthermore, we can utilise the properties of symmetric functions to verify the identity
    \begin{equation}
        (-1)^{\kappa - 1} \bm \partial_{\bm x}^{\bm k} \left[ \sum_{j=1}^N \frac{(\lambda - x_j)^{\kappa - 2}}{\prod_{\ell \neq j}(x_j - x_{\ell})} \Theta[\lambda - x_j] \right] = \bm \partial_{\bm x}^{\bm k} \left[ \sum_{j=1}^N \frac{(x_j - \lambda)^{\kappa - 2}}{\prod_{\ell \neq j}(x_j - x_{\ell})} \Theta[ x_j - \lambda ] \right] \ .
    \end{equation}
    A more general version of this equation, valid for rank-$R$ matrices, is derived in \ref{app:sym_funcs_rankR} and we obtain the above result by setting $R=1$ in Eq. \eqref{eq:app_sym_funcs_results}. This allows us to write
    \begin{equation}
        \mathcal{P}_{\bm z, \bm k}(\lambda) = \frac{\Gamma(\kappa)}{\Gamma(\kappa - 1)} \, \bm \partial_{\bm x}^{\bm k} \, \left[ \sum\limits_{j=1}^{N} \frac{\left( x_j - \lambda \right)^{\kappa-2}}{\prod\limits_{l\neq j}^N (x_j - x_l)}  \Theta\left[ x_j - \lambda \right] \right]_{\bm x = \bm z} \ ,
        \label{eq:pdf_lambda_form2}
    \end{equation}
    thus concluding our proof of Eq. \eqref{eq:pdf_lambda_GDRV}. Note that the two forms of the pdf given in Eqs. \eqref{eq:pdf_lambda_form1} and \eqref{eq:pdf_lambda_form2} are completely equivalent, with the choice of using one version or the other dependent upon certain specific contexts.

    One can now utilise Eq. \eqref{eq:pdf_lambda_form1} to obtain the non-negative integer moments ($m \geq 0$) of $\lambda$ as
    \begin{align}
        \expval{\lambda^m}_{\bm a} \equiv& \int_0^\infty \diff \lambda \, \lambda^m \, \mathcal{P}_{\bm z, \bm k}(\lambda) = \frac{\Gamma(\kappa)}{\Gamma(\kappa - 1)} \, \bm \partial_{\bm x}^{\bm k} \, \left[ \sum_{j=1}^N \frac{1}{\prod_{\ell \neq j}(x_j - x_\ell)} \int_0^{x_j} \diff \lambda \, \lambda^m \, (x_j - \lambda)^{\kappa - 2} \right]_{\bm x = \bm z} \nonumber\\
        =& \frac{\Gamma(\kappa) \, \Gamma(m+1)}{\Gamma(\kappa + m)} \, \bm \partial_{\bm x}^{\bm k} \, \left[ \sum_{j=1}^N \frac{ x_j^{\kappa + m - 1} }{\prod_{\ell \neq j}(x_j - x_\ell)}  \right]_{\bm x = \bm z} \ .
    \end{align}    
    Finally, by setting $m=0$, we can also demonstrate the normalisation of this pdf since 
    \begin{equation}
        \int_0^\infty \diff \lambda \, \mathcal{P}_{\bm z, \bm k}(\lambda) = \bm \partial_{\bm x}^{\bm k} \left[ \sum_{j=1}^N \frac{x_j^{\kappa-1}}{\prod_{\ell \neq j} (x_j - x_{\ell})}  \right]_{\bm x = \bm z} \ .
    \end{equation}
    As can be seen from Eq. \eqref{eq:S_m}, if $x_j$ has an exponent of $\kappa - 1$ then the summation over $j$, will in turn yield a summation over all combinations of $x_1^{i_1} x_2^{i_2} \cdots x_N^{i_N}$ such that $\sum_j i_j = \kappa - N$. Therefore upon the action of $\bm \partial_{\bm x}^{\bm k}$ and its $\kappa - N$ derivatives, the only term that would contribute would be proportional to $x_1^{k_1 - 1} x_2^{k_2 - 1} \cdots x_N^{k_N -1}$ and so
    \begin{equation}
        \bm \partial_{\bm x}^{\bm k} \left[ \sum_{j=1}^N \frac{x_j^{\kappa-1}}{\prod_{\ell \neq j} (x_j - x_{\ell})}  \right]_{\bm x = \bm z} = \left[\prod_{i=1}^N \frac{1}{\Gamma(k_i)} \frac{\partial^{k_i-1}}{ \partial x_i^{k_i-1} } x_i^{k_i - 1}\right]_{\bm x = \bm z} = 1 \ ,
        \label{eq:norm_arg}
    \end{equation}
    proving that the pdf is appropriately normalised.
\end{proof}

\subsection{Non-zero singular value of rank-$1$ NNEMs}
\label{sec:proofs_sig_M1}

We now move onto a proof of the statements of Theorem \ref{thm:sig2_M1}, which gives the integer moments of the non-zero squared singular value of the matrix $M_1$. In order to do this, we introduce the following Lemma which will be useful during this proof and in deriving further results.
\begin{lem} \label{lem:multinomial_sum}
    A multinomial sum over $N$ non-negative indices $m_1, m_2,\ldots,m_N$ with the constraint that $m_1+\cdots+m_N = m$, can be rewritten as the limit of a derivative in the following way
    \begin{align}
        \sum_{m_1+\cdots+m_N=m}^m& f(m_1, m_2, \ldots, m_N) = \nonumber \\
        &\frac{1}{m!} \lim_{w \to 0 } \frac{\diff^m}{\diff w^m} \left[ \sum_{m_1=0}^m w^{m_1} \sum_{m_2=0}^m w^{m_2}  \cdots \sum_{m_N=0}^m w^{m_N} f(m_1, m_2, \ldots, m_N) \right] \ .
        \label{eq:multi_sum_1}
    \end{align}
    Furthermore, if $f(m_1, m_2, \ldots, m_N)$ has a factorisable form, i.e. we can write $f(m_1, m_2, \ldots, m_N) = f_1(m_1) f_2(m_2) \ldots f_N(m_N)$, then it follows that 
    \begin{align}
        \sum_{m_1+\cdots+m_N=m}^m& \prod_{i=1}^N f_i(m_i) = \frac{1}{m!} \lim_{w \to 0 } \frac{\diff^m}{\diff w^m} \left[ \prod_{i=1}^N \left( \sum_{j=0}^m w^j f_{i}(j) \right) \right] \ .
        \label{eq:multi_sum_2}
    \end{align}
\end{lem}
\noindent
The proof of this Lemma goes as follows:
\begin{proof}
    We start by introducing the function
    \begin{equation}
        F(m) \equiv \sum_{m_1+\cdots+m_N=m}^m f(m_1, m_2, \ldots, m_N)  \ ,
    \end{equation}
    and immediately we note that the constrained sum can be rewritten using a Kronecker-delta, i.e.
    \begin{equation}
        F(m) = \sum_{m_1=0}^m \sum_{m_2=0}^m \cdots \sum_{m_N=0}^m f(m_1, m_2, \ldots, m_N) \delta_{\sum m_i, m}  \ .
    \end{equation}
    We then employ the following integral representation of the Kronecker delta
    \begin{equation}
        \delta_{m,n} = \frac{1}{2 \pi i} \oint_{|w|=1} \diff w \, w^{m-n-1} \ ,
        \label{eq:Kronecker_int_rep}
    \end{equation}
    which allows us to write that
    \begin{equation}
        F(m) = \frac{1}{2 \pi i} \oint_{|w|=1} \frac{\diff w}{w^{m+1}} \left[ \sum_{m_1=0}^m w^{m_1} \sum_{m_2=0}^m w^{m_2} \cdots \sum_{m_N=0}^m w^{m_N} f(m_1, m_2, \ldots, m_N) \right] \ .
    \end{equation}
    This can be evaluated through the use of Cauchy's residue theorem as
    \begin{equation}
        F(m) = \frac{1}{m!} \lim_{w \to 0} \frac{\diff^m}{\diff w^m} \left[ \sum_{m_1=0}^m w^{m_1} \sum_{m_2=0}^m w^{m_2} \cdots \sum_{m_N=0}^m w^{m_N} f(m_1, m_2, \ldots, m_N) \right] \ ,
    \end{equation}
    thus proving Eq. \eqref{eq:multi_sum_1}. One can now apply a factorisable form for the function that we are summing over, i.e. $f(m_1, m_2, \ldots, m_N) = f_1(m_1) f_2(m_2)\ldots f_N(m_N)$, and, upon realising that the summation can be rewritten using product notation, Eq. \eqref{eq:multi_sum_2} follows. Hence concluding our proof. 
\end{proof}
Note that this Lemma is useful as the sum over all integer partitions of $m$ of size $N$ is exponentially difficult to calculate, even for moderate $m$ and $N$. Employing this Lemma allows one to find a formula that is simpler and faster to evaluate using symbolic software (i.e. Mathematica). We now move onto proving the results related to the non-zero singular value of $M_1$ and begin by proving the general result of Eq. \eqref{eq:moms_sig2_GDRV} and then we specialise to the case of flat DRVs to obtain Eq. \eqref{eq:moms_sig2_FDRV}.

\begin{proof}
    The non-zero squared singular value of $M_1$ is given in Eq. \eqref{eq:lambda_svs_def} as $\sigma^2 = r^2 \bm a^T \bm a$, where $r^2$ is a constant. This implies that the associated moments of $\sigma^2$ can be written as $\expval{\sigma^{2m}}_{\bm a} = r^{2m} \expval{\varphi^m}_{\bm a}$, where
    \begin{equation}
        \varphi \equiv \bm a^T \bm a = \sum_{i=1}^N a_i^2 \ .
    \end{equation}
    Therefore, the problem of calculating the integer moments of the squared singular value coincides with obtaining the moments of the sum of the squares of DRVs. To this end, we introduce the function $f(m,t)$, such that $f(m,1) = \expval{\varphi^m}$, which is defined according to
    \begin{align}
        f_{\bm k}(m,t) \equiv& \frac{1}{B(\bm k)} \left[ \prod_{i=1}^N \int_0^\infty \diff a_i \, a_i^{k_i - 1}  \right] \left( \sum_{i=1}^N a_i^2 \right)^m \delta \left( \sum_{i=1}^N a_i - t \right) \nonumber \\
        =& \frac{1}{B(\bm k)}  \sum_{m_1 + \cdots + m_N = m}^m \frac{m!}{m_1! m_2! \cdots m_N!} \, I_{\bm k}(\bm m,t) \ , 
        \label{eq:f_pre_LT} 
    \end{align}
    where in the final line we have employed the multinomial theorem and introduced 
    \begin{equation}
        I_{\bm k}(\bm m, t) \equiv \left[ \prod_{i=1}^N \int_0^\infty \diff a_i \, a_i^{k_i + 2m_i - 1}  \right] \delta \left( \sum_{i=1}^N a_i - t \right)
    \end{equation}
    with $\bm m = (m_1, m_2, \ldots, m_N)^T$. At this point, we now perform a Laplace transform on $I_{\bm k}(\bm m, t)$ w.r.t. $t$ which yields
    \begin{align}
        \widetilde{I}_{\bm k}(\bm m,s) \equiv& \int_0^\infty \diff t \, e^{-st} I_{\bm k}( \bm m,t) \nonumber \\
        =& \prod_{i=1}^N \left[\int_0^\infty \diff a_i \, a_i^{k_i + 2m_i - 1} e^{-s a_i} \right] 
        = \frac{1 }{s^{\kappa + 2 m}} \prod_{i=1}^N \Gamma(k_i + 2m_i) \ , 
        \label{eq:varphi_pre_lemma}
    \end{align}
    then the corresponding ILT of $s\to t$ can be carried out to give
    \begin{equation}
        I_{\bm k}(\bm m, t) = \frac{t^{\kappa + 2m - 1}}{\Gamma(\kappa + 2m)} \prod_{i=1}^N \Gamma(k_i + 2m_i) \ ,
    \end{equation}
    where $\kappa = \sum_i k_i$ as previously. Inserting this result back into Eq. \eqref{eq:f_pre_LT}, after setting $t=1$, we see that
    \begin{equation}
        \expval{\varphi^m}_{\bm a} = \frac{1}{B(\bm k)} \frac{m!}{\Gamma(\kappa + 2m)} \chi_{\bm k}(m) \ ,
        \label{eq:mom_phi_chi_def}
    \end{equation}
    with
    \begin{equation}
        \chi_{\bm k}(m) \equiv \sum_{m_1 + \cdots + m_N = m}^m \left[ \prod_{i=1}^N \frac{\Gamma(2m_i + k_i)}{\Gamma(m_i + 1)} \right] \ .
    \end{equation}
    Through the use of Eq. \eqref{eq:multi_sum_2} of Lemma \ref{lem:multinomial_sum} we can see that this can be rewritten as
    \begin{align}
        \chi_{\bm k}(m) =&  \frac{1}{m!} \lim_{w \to 0} \frac{\diff^m}{\diff w^m} \left[ \prod_{i=1}^N \left(  \sum_{j=0}^m \, \frac{\Gamma(2j + k_i)}{\Gamma(j + 1)} \, w^j \right)\right] \ ,
        \label{eq:chi_varphi}
    \end{align}
    thus, inserting Eq. \eqref{eq:chi_varphi} into Eq. \eqref{eq:varphi_pre_lemma} we prove our result in Eq. \eqref{eq:moms_sig2_GDRV}.
    
    In order to prove Eq. \eqref{eq:moms_sig2_FDRV} for the integer moments of $\sigma^2$ when the DRVs are flat, we start from Eq. \eqref{eq:moms_sig2_GDRV}, which is valid for all positive integer choices of $k_i$. We first look to specialise this result for the case of symmetric DRVs where $k_1 = \cdots = k_N = k$, for all $i=1,\ldots , N$, and then at the end we set $k=1$. In this scenario, 
    \begin{equation}
        \expval{\varphi^m}_{\bm a} = \frac{1}{B(\bm k)} \frac{1}{\Gamma(Nk + 2m)} \lim_{w \to 0} \frac{\diff^m}{\diff w^m} \left[ \sum_{j=0}^m \, \frac{\Gamma(2j + k)}{\Gamma(j + 1)} \, w^j \right]^N \ , 
    \end{equation}
    which can be further analysed through the use of the Fa\`{a} di Bruno formula for the $m$-th derivative of the composition of two functions,
    \begin{equation}
        \frac{\diff^m}{\diff z^m} \phi(g(z)) = \sum_{\ell=1}^m \phi^{(\ell)}(g(z)) \, B_{m,\ell} \Big(g^{(1)}(z), g^{(2)}(z), \cdots , g^{(m-\ell+1)}(z) \Big) \ ,
    \end{equation}
    where $\phi^{(\ell)}(x)$ and $g^{(\ell)}(x)$ are the $\ell$-th derivatives of $\phi$ and $g$ evaluated at $x$ respectively and $B_{m,\ell}(\cdot)$ are Bell polynomials \cite{Wang09}. In our case, we have that 
    \begin{equation}
        \phi(x) = x^N \hspace{2cm} \text{and} \hspace{2cm} g(z) = \sum_{j=0}^m f(j,k)z^j\ ,
    \end{equation}
    with $f(j,k) \equiv \Gamma(2j+k)/\Gamma(j+1)$ and derivatives
    \begin{equation}
        \phi^{(\ell)}(g(z))\Bigg|_{z=0}= \Theta[N-\ell] \frac{N!}{(N-\ell)!} \left[ \sum_{j=0}^m f(j,k) \, z^j \right]^{N-\ell}\Bigg|_{z=0} = \Theta[N-\ell] \, \Gamma(k)^{N-\ell} \, \frac{N!}{(N-\ell)!}  \ ,
    \end{equation}
    since $f(0,k)=\Gamma(k)$. Additionally, it is easy to see that the $\ell$-th derivative of $g(z)$ at $z=0$ is given by $g^{(l)}(z=0) = \ell! f(\ell,k)\Theta[m-\ell]$. Combining the above equations, we can see that 
    \begin{align}
        \expval{\varphi^m}_{\bm a} = \frac{1}{B(\bm k)} &\frac{1}{\Gamma(Nk + 2m)} \sum_{\ell=1}^{\min(N,m)} \frac{N!}{(N - \ell)!} \, \Gamma(k)^{N-\ell} \nonumber  \\
        &\times B_{m,\ell} \Big( 1! f(1,k), 2! f(2,k), \ldots , (m-\ell+1)! f(m-\ell+1,k)  \Big) 
    \end{align}
    and for the case of $k=1$ we can easily obtain Eq. \eqref{eq:moms_sig2_GDRV}, hereby concluding the proof. 
\end{proof}

\section{Proofs of results for rank-2 NNEMs}
\label{sec:rank2_proofs}

Firstly, in the setting of our rank-2 results, we begin by verifying the statement of Proposition \ref{prop:jpdf_alpha_beta}, for the jpdf  $\overline{\mathcal{P}}_{\bm x, \bm y, \bm k}(\alpha, \beta)$ of the random objects $\alpha$ and $\beta$.  For ease of notation in this proof, we suppress the use of the overbar and simply write $\mathcal{P}_{\bm x, \bm y, \bm k}(\alpha, \beta)$. Once this task is complete, in Section \ref{sec:proofs_lam_M2} we then prove the results outlined in Theorem \ref{thm:lam_M2} which refer to the non-zero eigenvalues of $M_2$. Then, in Section \ref{sec:proofs_sig_M2}, we prove the results of Theorem \ref{thm:sig2_M2}, which describe statistical properties of the squared singular values of $M_2$. 

\subsection{Jpdf of $\alpha$ \& $\beta$}
\label{sec:proof_jpdf_alpha_beta}

In what follows we prove the result given in Eq. \eqref{eq:jpdf_alpha_beta_N} of Proposition \ref{prop:jpdf_alpha_beta}, which outlines a result for the jpdf of $\alpha \equiv \bm a^T \bm x$ and $\beta \equiv \bm a^T \bm y$, for a general Dirichlet random vector $\bm a$ and two different fixed non-negative entry vectors $\bm x$ and $\bm y$.

\begin{proof}
    We start from the definition of the required jpdf which reads
    \begin{align}
        {\mathcal{P}}_{\bm x, \bm y, \bm k}(\alpha, \beta, t) \equiv \frac{1}{B(\bm k)} \int \diff \bm a \, &\left( \prod_{i=1}^N a_i^{k_i-1} \right)  \delta\left( \sum_{i=1}^{N} a_i - t \right) \nonumber \\
        &\times \delta\left(\alpha - \sum_{i=1}^N a_i x_i \right) \delta\left(\beta - \sum_{i=1}^N a_i y_i \right) \ , 
        \label{eq:pdf_alpha_beta_t_preLT}
    \end{align}
    where, once again, we have introduced an auxiliary variable $t$ for the linear sum of the DRVs, such that ${\mathcal{P}}_{\bm x, \bm y, \bm k}(\alpha, \beta) \equiv {\mathcal{P}}_{\bm x, \bm y, \bm k}(\alpha, \beta, 1)$. At this point we perform a triple Laplace transform w.r.t. the variables $\alpha$, $\beta$ and $t$ and then trivially carry out the integration over the vector $\bm a$, such that
    \begin{align}
        \widetilde{\mathcal{P}}^{(3)}_{\bm x, \bm y, \bm k}&(A, B, s) \equiv \int_0^\infty \diff \alpha \, e^{- A\alpha} \int_0^\infty \diff \beta \, e^{- B \beta}  \int_0^\infty \diff t \, e^{- s t} \, \mathcal{P}_{\bm x, \bm y, \bm k}(\alpha, \beta, t) \nonumber \\
        =& \frac{1}{B(\bm k)}\int \diff \bm a \, \left( \prod_{i=1}^N a_i^{k_i-1} \right) \, \exp\left[ - s \sum_{i=1}^{N} a_i  - A \sum_{i=1}^N a_i x_i - B \sum_{i=1}^N a_i y_i \right] \nonumber \\
        =& \Gamma(\kappa) \prod_{i=1}^N \frac{1}{(s + A x_i + B y_i)^{k_i}} \ .
        \label{eq:p3_pre_derivs}
    \end{align}
   Using the identity
    \begin{equation}
        \frac{1}{(s + A x_i + B y_i)^{k_i}} = \frac{(-1)^{k_i - 1}}{\Gamma(k_i) A^{k_i - 1}} \frac{\partial^{k_i - 1}}{\partial v_i^{k_i - 1}} \left( \frac{1}{s + A v_i + B y_i} \right)\Bigg|_{v_i = x_i} \ ,
    \end{equation}
    we can re-express Eq. \eqref{eq:p3_pre_derivs} in terms of a derivative over the vector $\bm v = (v_1, v_2,\ldots,v_N)^T$ as
    \begin{equation}
        \widetilde{\mathcal{P}}^{(3)}_{\bm x, \bm y, \bm k}(A, B, s) = \frac{\Gamma(\kappa)(-1)^{\kappa - N}}{A^{\kappa - N}} \bm \partial_{\bm v}^{\bm k} \left[ \prod_{i=1}^N \frac{1}{s + A v_i + By_i} \right]_{\bm v = \bm x} \ ,
    \end{equation}
    where $\bm \partial_{\bm v}^{\bm k}$ is defined according to Eq. \eqref{eq:def_partials}. Note that we could easily have replaced $\bm y$ with $\bm v$ and followed the same process of differentiation - the overall result is independent of this choice. The main computational difficulty that remains at this point is in performing the three Laplace inversions. Introducing $\zeta_i \equiv A v_i + B y_i$, we may use the partial fraction expansion
    \begin{equation}
        \prod_{i=1}^N \frac{1}{s + \zeta_i} = \sum_{i=1}^N \frac{(-1)^{N-1}}{\prod\limits_{\ell \neq i}(\zeta_i - \zeta_{\ell})} \frac{1}{ s + \zeta_i } \ ,
        \label{eq:3Laplace_invs}
    \end{equation}
    which allows us to perform the inversion w.r.t. $s$, after setting $t=1$ we see that
    \begin{align}
        \widetilde{\mathcal{P}}^{(2)}_{\bm x, \bm y, \bm k}(A, B) \equiv& \mathcal{L}^{-1}_{s\to t} \left[ \widetilde{\mathcal{P}}^{(3)}_{\bm x, \bm y, \bm k}(A, B, s) \right]\Big|_{t=1} \nonumber \\
        =& \frac{\Gamma(\kappa) (-1)^{\kappa-1}}{A^{\kappa - N}} \, \bm \partial_{\bm v}^{\bm k} \, \left[ \sum_{i=1}^N \frac{e^{-Av_i - B y_i}}{ \prod\limits_{\ell = 1 , \ell \neq i}^N \left( A V_{i\ell} + B Y_{i\ell} \right)} \right]_{\bm v = \bm x} \ ,
    \end{align}
    where $V_{i \ell} \equiv v_i - v_{\ell}$ and $Y_{i \ell} \equiv y_i - y_{\ell}$. To proceed further from here we again use partial fractions by employing the identity
    \begin{equation}
        \prod\limits_{\ell = 1, \ell \neq i}^N\frac{1}{\Big( A V_{i\ell} +  B Y_{i\ell} \Big)} = \sum_{j=1, j \neq i}^N \frac{Y_{ij}^{N-2}}{\prod\limits^N_{\ell = 1, \ell \neq i \neq j} \Big( Y_{ij} V_{i\ell} - Y_{i\ell} V_{ij} \Big) } \frac{1}{A^{N-2} \Big( A V_{ij} +  B Y_{ij} \Big)} \ ,
    \end{equation}
    which can be derived through induction. Therefore, in order to obtain the required jpdf, one must now perform a double ILT such that
    \begin{align}
        \mathcal{P}_{\bm x, \bm y, \bm k}(\alpha, \beta) = \Gamma(\kappa) (-1)^{\kappa-1} \, \bm \partial_{\bm v}^{\bm k} \, \Bigg[  \sum_{i=1}^N \sum_{j=1, j \neq i}^N & \frac{Y_{ij}^{N-2}}{\prod\limits^N_{\ell = 1, \ell \neq i \neq j} \Big( Y_{ij} V_{i\ell} - Y_{i\ell} V_{ij} \Big)  } \nonumber \\
        & \times \mathcal{L}^{-1}_{ A\to \alpha, B \to \beta} \left[\widetilde{\mathcal{F}}^{(2)}_{\bm v, \bm y, \kappa} (A,B;i,j) \right] \Bigg]_{\bm v = \bm x} \ , 
    \end{align}
    where we have introduced
    \begin{equation}
        \widetilde{\mathcal{F}}^{(2)}_{\bm v, \bm y, \kappa} (A,B;i,j) \equiv \frac{e^{- A v_i - B y_i}}{A^{\kappa-2} \Big( A V_{ij} +  B Y_{ij} \Big)} \ .
    \end{equation}
    Starting with the special case of $\kappa=2$, and carrying out the inversion w.r.t. $B$ first, we see that
    \begin{align}
        \widetilde{\mathcal{F}}^{(1)}_{\bm v, \bm y, 2} (A:\beta;i,j) \equiv& \mathcal{L}^{-1}_{B \to \beta} \left[ \mathcal{F}^{(2)}_{\bm v, \bm y, 2} (A,B;i,j) \right] \nonumber \\
        =& \frac{1}{Y_{ij}} \exp\left[ - A\left( v_i + \frac{V_{ij}}{Y_{ij}}(\beta - y_i) \right) \right]\Theta[\beta - y_i] 
    \end{align}
    and then inverting w.r.t. $A$ we get
    \begin{align}
        \mathcal{F}_{\bm v, \bm y, 2} (\alpha, \beta;i,j) \equiv& \mathcal{L}^{-1}_{A \to \alpha} \left[ \mathcal{F}^{(1)}_{\bm v, \bm y, 2} (A:\beta;i,j) \right] = \frac{1}{Y_{ij}} \delta\left( \alpha - v_i - \frac{V_{ij}}{Y_{ij}}(\beta - y_i) \right) \Theta[\beta - y_i] \ . 
    \end{align}
    Thus for $\kappa=2$, we can write the jpdf of $\alpha$ and $\beta$ exactly as written in the first case of Eq. \eqref{eq:jpdf_alpha_beta_N} - as the only way that $\kappa$ can equal 2 is for a $2\times 2$ matrix with $k_1 = k_2 =1$, hence there are no derivatives to take. Note that in order to remove the summation we have used the property that, for any pair of $i$ and $j$, $V_{ij} = -V_{ji}$ and $Y_{ij} = -Y_{ji}$, which allows us to see that
    \begin{equation}
         \alpha - v_i - \frac{V_{ij}}{Y_{ij}}(\beta - y_i) = \alpha - v_j - \frac{V_{ji}}{Y_{ji}}(\beta - y_j) \ .
         \label{eq:Vij_Yij_exchange}
    \end{equation}
    Now, considering the more general case of $\kappa \geq 3$, we carry out the Laplace inversion w.r.t. $B$ and find that
    \begin{align}
        \widetilde{\mathcal{F}}^{(1)}_{\bm v, \bm y, \kappa} (A:\beta;i,j) \equiv& \mathcal{L}^{-1}_{B \to \beta}\left[ \widetilde{\mathcal{F}}^{(2)}_{\bm v, \bm y, \kappa} (A,B;i,j) \right] \nonumber \\
        =& \frac{1}{Y_{ij}} \frac{1}{A^{\kappa-2}} \exp\left[- A\left( v_i + \frac{V_{ij}}{Y_{ij}}(\beta - y_i)\right) \right] \Theta[\beta - y_i] \ .
    \end{align}
    This can be further inverted w.r.t. $A$, yielding
    \begin{align}
        &\mathcal{F}_{\bm v, \bm y, \kappa} (\alpha,\beta;i,j) \equiv \mathcal{L}^{-1}_{A \to \alpha} \left[ \widetilde{\mathcal{F}}^{(1)}_{\bm v, \bm y, \kappa} (A:\beta;i,j) \right] \nonumber \\
        &= \frac{1}{\Gamma(\kappa-2)} \frac{1}{Y_{ij}} \left( \alpha - v_i - \frac{V_{ij}}{Y_{ij}}(\beta - y_i) \right)^{\kappa-3} \Theta\left[ \alpha - v_i - \frac{V_{ij}}{Y_{ij}}(\beta - y_i)  \right]\Theta[\beta - y_i] \ .
    \end{align}
    Thus, we can write that
     \begin{align}
        \mathcal{P}_{\bm x, \bm y, \bm k}&(\alpha, \beta) = \frac{\Gamma(\kappa)(-1)^{\kappa-1}}{\Gamma(\kappa-2)} \, \bm \partial_{\bm v}^{\bm k} \, \Bigg[ \sum_{i=1}^N \sum_{j=1 , j \neq i}^N \frac{Y_{ij}^{N-3}}{\prod\limits^N_{\ell = 1, \ell \neq i \neq j} \Big( Y_{ij} V_{i\ell} - Y_{i\ell} V_{ij} \Big) } \nonumber \\
        &\times \left( \alpha - v_i - \frac{V_{ij}}{Y_{ij}}(\beta - y_i) \right)^{\kappa-3} \Theta[\beta - y_i] \Theta\left[ \alpha - v_i - \frac{V_{ij}}{Y_{ij}}(\beta - y_i) \right] \Bigg]_{\bm v = \bm x} \label{eq:jpdf_alpha_beta_form1}\ ,
    \end{align}
    which can, eventually, be manipulated into the form of Eq. \eqref{eq:jpdf_alpha_beta_N}. To do this we first note that 
    \begin{equation}
        Y_{ij} V_{i\ell} - Y_{i\ell} V_{ij} = \Delta_{j \ell} + \Delta_{ij} + \Delta_{\ell i} \ ,
    \end{equation}
    with $\Delta_{ij} \equiv y_i v_j - y_j v_i$, which makes it easy to see that
    \begin{equation}
        W_{ij} = \frac{Y_{ij}^{N-3}}{\prod\limits^N_{\ell = 1, \ell \neq i \neq j} \Big( \Delta_{j \ell} + \Delta_{ij} + \Delta_{\ell i} \Big) } = \frac{(-1)^{N-3} \, Y_{ji}^{N-3}}{(-1)^{N-2}\prod\limits^N_{\ell = 1, \ell \neq i \neq j} \Big(  \Delta_{\ell j} + \Delta_{ji} + \Delta_{i \ell} \Big) } = -W_{ji} \ .
    \end{equation}    
    We can use the above in conjunction with Eq. \eqref{eq:Vij_Yij_exchange} to re-express the sum over $j \neq i$ in Eq. \eqref{eq:jpdf_alpha_beta_form1} as a sum over all pairs of $j$ and $i$, which directly leads to Eq. \eqref{eq:jpdf_alpha_beta_N} - concluding our proof.
\end{proof}

\subsection{Non-zero eigenvalues of rank-$2$ NNEMs}
\label{sec:proofs_lam_M2}

With this result for the jpdf in place, we are now in a position to prove Theorem \ref{thm:lam_M2}. 

\begin{proof}
    We start by utilising the jpdf of $\alpha$ and $\beta$, denoted by $\overline{\mathcal{P}}_{\bm x, \bm y, \bm k}(\alpha, \beta)$ and given in Eq. \eqref{eq:jpdf_alpha_beta_N}, to write down an integral expression for the jpdf of the sum and product of the non-zero eigenvalues
    \begin{align}
        \mathcal{P}_{\bm x, \bm y, \bm h, \bm k}(T, D) \equiv \int_0^\infty \diff \alpha \int_0^\infty \diff \beta \int_0^\infty \diff \gamma \int_0^\infty \diff \zeta \, \overline{\mathcal{P}}_{\bm x, \bm y, \bm h}(\alpha, \beta) \, \overline{\mathcal{P}}_{\bm x, \bm y, \bm k}(\gamma, \zeta) \nonumber \\
        \times \delta\Big( T - \alpha - \zeta \Big) \, \delta\Big( D - \alpha \zeta + \beta \gamma \Big) \ ,
    \end{align}
    where we have utilised the definitions of $T$ and $D$ from Eqs. \eqref{eq:T_def} and \eqref{eq:D_def} respectively. Performing the integral over $\zeta$ yields
    \begin{align}
        \mathcal{P}_{\bm x, \bm y, \bm h, \bm k}(T, D) = \int_0^\infty \diff \alpha \int_0^\infty \diff \beta \int_0^\infty \diff \gamma \, \overline{\mathcal{P}}_{\bm x, \bm y, \bm h}(\alpha, \beta) \, \overline{\mathcal{P}}_{\bm x, \bm y, \bm k}(\gamma, T - \alpha) \nonumber \\
        \times \delta\Big( D - \alpha (T - \alpha) + \beta \gamma \Big) \, \Theta[T - \alpha]\ , 
    \end{align}
    then after manipulating the remaining $\delta$-function we can evaluate the integral over $\gamma$ and find that
    \begin{align}
        \mathcal{P}_{\bm x, \bm y, \bm h, \bm k}(T, D) = \int_0^\infty \diff \alpha \int_0^\infty \frac{\diff \beta}{\beta} \, \overline{\mathcal{P}}_{\bm x, \bm y, \bm h}(\alpha, \beta) \, \overline{\mathcal{P}}_{\bm x, \bm y, \bm k}\left(\frac{\alpha (T - \alpha) - D}{\beta}, T - \alpha \right) \, \nonumber \\ 
        \times \Theta[T - \alpha] \, \Theta\left[ \frac{\alpha (T - \alpha) - D}{\beta} \right] \ , 
    \end{align}
    as written in Eq. \eqref{eq:jpdf_T_D}. Furthermore, noting that $T \equiv \lambda_1 + \lambda_2$ and $D \equiv \lambda_1\lambda_2$, we can use the jpdf of the sum and product of the eigenvalues to obtain an expression for the jpdf of the non-zero eigenvalues themselves, which reads
    \begin{align}
        \mathcal{P}_{\bm x, \bm y, \bm h, \bm k}^{(\lambda_1, \lambda_2)}(\lambda_1, \lambda_2) \equiv& \int_0^\infty \diff T \int_{-\infty}^\infty \diff D \, \mathcal{P}_{\bm x, \bm y, \bm h, \bm k}(T, D) \, \delta\Big( T - \lambda_1 - \lambda_2 \Big) \, \delta\Big( D - \lambda_1 \lambda_2 \Big) \nonumber \\
        =& \mathcal{P}_{\bm x, \bm y, \bm h, \bm k}\Big(\lambda_1 + \lambda_2, \lambda_1 \lambda_2 \Big) \Theta\Big[\lambda_1 + \lambda_2 \Big] \ .
    \end{align}
    We defer the proofs of Eqs. \eqref{eq:pdf_T_M2} and \eqref{eq:moms_T_M2} to Section \ref{sec:rankR_proofs} because these results can be easily obtained from Eqs. \eqref{eq:pdf_T_MR} and \eqref{eq:moms_T_MR} by setting $R=2$ and rewriting $\bm z_1 \to \bm x$, $\bm z_2 \to \bm y$, $\bm k_1 \to \bm h$ and $\bm k_2 \to \bm k$. We now move onto proving the result from Eq. \eqref{eq:moms_D_GDRV} for the moments of the product of the non-zero eigenvalues $D$. It is helpful to begin by introducing the object
    \begin{align}
        \Lambda_{\bm x, \bm y, \bm h, \bm k}(m, t_1, t_2) \equiv& \frac{1}{B(\bm h) B(\bm k)} \int \diff \bm a  \left( \prod_{i=1}^N a_i^{h_i - 1} \right) \int \diff \bm b \left( \prod_{i=1}^N b_i^{k_i - 1} \right) \delta\left( \sum_{i=1}^N a_i - t_1 \right) \nonumber\\
        & \times \delta\left( \sum_{i=1}^N b_i - t_2 \right) \bigg( (\bm a^T \bm x) (\bm b^T \bm y) - (\bm a^T \bm y) (\bm b^T \bm x) \bigg)^m  \ ,
    \end{align}
    with $t_1$ and $t_2$ being used as auxiliary variables for the linear sum of the DRVs - as usual these will be set to one later. Additionally, we define $\Lambda_{\bm x, \bm y, \bm h, \bm k}(m) \equiv \Lambda_{\bm x, \bm y, \bm h, \bm k}(m,1,1) = \expval{D^m}_{\bm a, \bm b}$. Employing the multinomial theorem we see that
    \begin{align}
        \Lambda_{\bm x, \bm y, \bm h, \bm k}(m, t_1, t_2) =& \frac{1}{B(\bm h) B(\bm k)} \sum_{m_1 + m_2 = m}^m \frac{m!(-1)^{m_2}}{m_1! m_2!} \int \diff \bm a  \left( \prod_{i=1}^N a_i^{h_i - 1} \right) \int \diff \bm b \left( \prod_{i=1}^N b_i^{k_i - 1} \right) \nonumber \\
        & \times \delta\left( \sum_{i=1}^N a_i - t_1 \right) \delta\left( \sum_{i=1}^N b_i - t_2 \right)  \nonumber \\
        &\times \left( \sum_{i=1}^N a_i x_i \right)^{m_1} \left( \sum_{i=1}^N b_i y_i \right)^{m_1} \left( \sum_{i=1}^N a_i y_i \right)^{m_2} \left( \sum_{i=1}^N b_i x_i \right)^{m_2}  \ ,
    \end{align}
    then, after a further four applications of the multinomial theorem, we see that 
    \begin{align}
        \Lambda_{\bm x, \bm y, \bm h, \bm k}(m, t_1, t_2) =& \frac{m!}{B(\bm h) B(\bm k)} \sum_{m_1 + m_2 = m}^m m_1! \, m_2!\, (-1)^{m_2}  \sum_{n_1+\cdots +n_N = m_1}^{m_1} \sum_{p_1+\cdots +p_N = m_1}^{m_1}  \nonumber \\
        & \times \sum_{q_1+\cdots +q_N = m_2}^{m_2} \sum_{r_1+\cdots +r_N = m_2}^{m_2} \prod_{i=1}^N \frac{x_i^{n_i + r_i} y_i^{p_i+q_i}}{n_i! p_i! q_i! r_i!} \int \diff \bm a  \,  \delta\left( \sum_{i=1}^N a_i - t_1 \right) \nonumber  \\
        & \times \int \diff \bm b \, \delta\left( \sum_{i=1}^N b_i - t_2 \right) \left( \prod_{i=1}^N a_i^{n_i + q_i + h_i - 1} \, b_i^{p_i + r_i + k_i - 1} \right)\ . \label{eq:refer_sums}
    \end{align}
    To evaluate the integrals over $\bm a$ and $\bm b$, we introduce the function
    \begin{align}
        \Omega_{\bm h, \bm k}(t_1, t_2 ; \bm n, \bm p, \bm q, \bm r) \equiv& \int \diff \bm a \, \delta\left( \sum_{i=1}^N a_i - t_1 \right) \left( \prod_{i=1}^N a_i^{n_i + q_i + h_i - 1} \right) \nonumber \\
        &\times \int \diff \bm b \, \delta\left( \sum_{i=1}^N b_i - t_2 \right) \left( \prod_{i=1}^N b_i^{p_i + r_i + k_i - 1} \right) \ ,
        \label{eq:int_steps_start}
    \end{align}
    with $\bm n =(n_1, n_2, \ldots, n_N)^T$ and analogously for $\bm p$, $\bm q$ and $\bm r$. We then perform a double Laplace transform on $t_1$ and $t_2$, which allows us to easily carry out the integrals over $\bm a$ and $\bm b$, such that
    \begin{align}
        \widetilde{\Omega}_{\bm h, \bm k}^{(2)}&(s_1, s_2 ; \bm n, \bm p, \bm q, \bm r) \equiv \int_0^\infty \diff t_1 \, e^{-s_1 t_1} \int_0^\infty \diff t_2 \, e^{-s_2 t_2} \, \Omega_{\bm h, \bm k}(t_1, t_2 ; \bm n, \bm p, \bm q, \bm r) \nonumber \\
        =& \prod_{i=1}^N \int_0^\infty \diff a_i \, a_i^{n_i + q_i + h_i - 1} e^{-s_1 a_i}  \, \int_0^\infty \diff b_i \, b_i^{p_i + r_i + k_i - 1} e^{-s_2 b_i}  \nonumber \\
        =& \prod_{i=1}^N\left( \frac{\Gamma(n_i + q_i + h_i)}{s_1^{n_i+q_i+h_i}} \, \frac{\Gamma(p_i + r_i + k_i)}{s_2^{p_i+r_i+k_i}}  \right) = \frac{\prod\limits_{i=1}^N \Gamma(n_i + q_i + h_i) \, \Gamma(p_i + r_i + k_i) }{s_1^{m_1 + m_2 + \eta} \, s_2^{m_1 + m_2 + \kappa}} \ ,
    \end{align}
    where $\eta = \sum_i h_i$ and $\kappa = \sum_i k_i$. Note as well that we have utilised the constraints on the sums in Eq. \eqref{eq:refer_sums} to say that $\sum_i n_i = \sum_i p_i = m_1$ and $\sum_i q_i = \sum_i r_i = m_2$. At this point, we now perform the associated Laplace inversions and, after setting $t_1 = t_2 = 1$,  we obtain  
    \begin{equation}
        \Omega_{\bm h, \bm k}( 1, 1 ; \bm n, \bm p, \bm q, \bm r) = \frac{\prod\limits_{i=1}^N \Gamma(n_i + q_i + h_i) \, \Gamma(p_i + r_i + k_i) }{\Gamma(m_1 + m_2 + \eta) \, \Gamma(m_1 + m_2 + \kappa)} \ ,
        \label{eq:int_steps_finish}
    \end{equation}
    which yields
    \begingroup
    \allowdisplaybreaks
    \begin{align}
        \Lambda_{\bm x, \bm y, \bm h, \bm k}(m) = &\frac{m!}{B(\bm h) B(\bm k)} \sum_{m_1 + m_2 = m}^m \frac{ (-1)^{m_2} \, m_1! \, m_2!}{\Gamma(m_1 + m_2 + \eta) \, \Gamma(m_1 + m_2 + \kappa)} \nonumber \\
        &\times \sum_{n_1+\cdots +n_N = m_1}^{m_1} \prod_{i=1}^N \frac{x_i^{n_i}}{n_i!} \sum_{p_1+\cdots +p_N = m_1}^{m_1} \prod_{i=1}^N \frac{y_i^{p_i}}{p_i!} \sum_{q_1+\cdots +q_N = m_2}^{m_2} \prod_{i=1}^N \frac{y_i^{q_i} \Gamma(n_i + q_i + h_i)}{q_i!} \nonumber \\
        & \times \sum_{r_1+\cdots +r_N = m_2}^{m_2} \prod_{i=1}^N \frac{ x_i^{r_i} \Gamma(p_i + r_i + k_i) }{ r_i!} \ .
    \end{align}
    \endgroup
    We now introduce and evaluate the object
    \begin{align}
        Q_{\bm x, \bm k}(m_2, \bm p) \equiv& \sum_{r_1+\cdots +r_N = m_2}^{m_2} \prod_{i=1}^N \frac{ x_i^{r_i} \Gamma(p_i + r_i + k_i) }{ r_i!} \nonumber \\
        =& \frac{1}{m_2!} \lim_{w \to 0} \frac{\diff^{m_2}}{\diff w^{m_2}} \left[ \prod_{i=1}^N \left( \sum_{j=0}^{m_2} \frac{\Gamma(p_i + k_i + j) }{j!} (wx_i)^j \right) \right] \ ,
    \end{align}
    where in order to achieve the second equality we have employed Eq. \eqref{eq:multi_sum_2} of Lemma \ref{lem:multinomial_sum}. This object allows us to write that
    \begin{align}
        \Lambda_{\bm x, \bm y, \bm h, \bm k}&(m) = \frac{m!}{B(\bm h) B(\bm k)} \sum_{m_1 + m_2 = m}^m \frac{(-1)^{m_2} \, m_1! \, m_2!}{\Gamma(m_1 + m_2 + \eta) \, \Gamma(m_1 + m_2 + \kappa)} \nonumber  \\
        \times & \sum_{n_1+\cdots +n_N = m_1}^{m_1} \left( \prod_{i=1}^N \frac{x_i^{n_i}}{n_i!} \right) \, Q_{\bm y, \bm h}(m_2, \bm n) \sum_{p_1 + \cdots +p_N = m_1}^{m_1} \left( \prod_{i=1}^N \frac{y_i^{p_i}}{p_i!} \right)\, Q_{\bm x, \bm k}(m_2, \bm p) 
    \end{align}
    and so, we consider the function
    \begin{align}
        P_{\bm x, \bm y, \bm h}(m_1, m_2) \equiv& \sum_{n_1+\cdots +n_N = m_1}^{m_1} \left(\prod_{i=1}^N \frac{x_i^{n_i}}{n_i!} \right) \, Q_{\bm y, \bm h}(m_2, \bm n) \nonumber \\
        =& \frac{1}{m_1!} \lim_{w \to 0} \frac{\diff^{m_1}}{\diff w^{m_1}} \left[ \sum_{n_1=0}^{m_1} \frac{(wx_1)^{n_1}}{n_1!} \cdots \sum_{n_N=0}^{m_1} \frac{(wx_N)^{n_N}}{n_N!} \, Q_{\bm y, \bm h}(m_2, \bm n) \right] \ ,
    \end{align}
    where we have used Eq. \eqref{eq:multi_sum_1} of Lemma \ref{lem:multinomial_sum} to achieve the second equality. This then means that we can write
    \begin{align}
        \Lambda_{\bm x, \bm y, \bm h, \bm k}&(m) = \frac{m!}{B(\bm h) B(\bm k)} \sum_{m_1 = 0}^m \sum_{m_2 = 0}^m \frac{ (-1)^{m_2} \, m_1! \, m_2!}{\Gamma(m_1 + m_2 + \eta) \, \Gamma(m_1 + m_2 + \kappa)} \nonumber \\
        & \hspace{3cm }\times P_{\bm x, \bm y, \bm h}(m_1, m_2) \, P_{\bm y, \bm x, \bm k}(m_1, m_2) \delta_{m_1+m_2,m} \nonumber \\
        =& \frac{m!}{B(\bm h) B(\bm k)} \sum_{n = 0}^m \frac{ (-1)^{m-n} \, n! \, (m-n)! }{\Gamma(m + \eta) \Gamma(m + \kappa)} \, P_{\bm x, \bm y, \bm h}(n, m - n) \, P_{\bm y, \bm x, \bm k}(n, m - n) \ ,
    \end{align}
    which concludes our proof of Eq. \eqref{eq:moms_D_GDRV} and consequently also concludes the proof of our results in Theorem \ref{thm:lam_M2}.    
\end{proof}

\subsection{Non-zero singular values of rank-$2$ NNEMs}
\label{sec:proofs_sig_M2}

We now move onto the proof of our results for the non-zero squared singular values of $M_2$ from Theorem \ref{thm:sig2_M2}, starting with the integer moments of their sum in Eq. \eqref{eq:moms_tau_M2} and then moving onto the integer moments of their product in Eq. \eqref{eq:moms_Delta_ab}.

\begin{proof}
    Firstly, we begin by rewriting the sum of the squared singular values, denoted by $\tau$ and defined in Eq. \eqref{eq:tau_def}, as
    \begin{equation}
        \tau = r_{xx} \, \bm a^T \bm a + r_{yy} \, \bm b^T \bm b + 2r_{xy} \, \bm a^T \bm b \ ,
    \end{equation}
    where we have introduced $r_{xx} \equiv \bm x^T \bm x$, $r_{yy} \equiv \bm y^T \bm y$ and $r_{xy} \equiv \bm x^T \bm y$. Additionally, since $\bm a$ and $\bm b$ are general Dirichlet random vectors with Dirichlet parameters given by the entries of the vectors $\bm h$ and $\bm k$ respectively, we can define
    \begin{align}
        \Phi_{\bm x, \bm y, \bm h, \bm k}(m, t_1, t_2) \equiv& \frac{1}{B(\bm h) B(\bm k)} \int \diff \bm a \, \left( \prod_{i=1}^N a_i^{h_i - 1} \right) \, \int \diff \bm b \, \left( \prod_{i=1}^N b_i^{k_i - 1} \right) \, \delta \left( \sum_{i=1}^N a_i - t_1 \right) \nonumber \\
        & \times \delta \left( \sum_{i=1}^N b_i - t_2 \right) \bigg( r_{xx} \, \bm a^T \bm a + r_{yy} \, \bm b^T \bm b + 2r_{xy} \, \bm a^T \bm b \bigg)^m \ , 
        \label{eq:tau_moms_start}
    \end{align}
    with auxiliary variables $t_1$ and $t_2$ used for the sums of the DRVs, such that $\Phi_{\bm x, \bm y, \bm h, \bm k}(m) \equiv \Phi_{\bm x, \bm y, \bm h, \bm k}(m,1,1) = \expval{\tau^m}_{\bm a, \bm b}$. Proceeding in a similar way to the proof of Eq. \eqref{eq:moms_sig2_GDRV}, we utilise the multinomial theorem to expand the final bracket of Eq. \eqref{eq:tau_moms_start} yielding
    \begin{align}
        \Phi_{\bm x, \bm y , \bm h, \bm k}(m, t_1, t_2) =& \frac{1}{B(\bm h) B(\bm k)} \sum_{m_1 + m_2 + m_3 = m}^m \frac{m!}{m_1! \, m_2! \, m_3!} \int \diff \bm a \, \left( \prod_{i=1}^N a_i^{h_i - 1} \right) \nonumber \\ 
        & \times \int \diff \bm b \, \left( \prod_{i=1}^N b_i^{k_i - 1} \right) \, \delta \left( \sum_{i=1}^N a_i - t_1 \right) \delta \left( \sum_{i=1}^N b_i - t_2 \right) \nonumber \\
        & \times r_{xx}^{m_1} \left( \sum_{i=1}^N a_i^2 \right)^{m_1} r_{yy}^{m_2} \left( \sum_{i=1}^N b_i^2 \right)^{m_2} (2r_{xy})^{m_3} \left( \sum_{i=1}^N a_i b_i \right)^{m_3}  \ .
    \end{align}
    Utilising the multinomial theorem a further three times we see that 
    \begingroup
    \allowdisplaybreaks
    \begin{align}
        \Phi_{\bm x, \bm y, \bm h, \bm k}(m, t_1, t_2) &= \frac{m!}{B(\bm h) B(\bm k)} \sum_{m_1 + m_2 + m_3 = m}^m  r_{xx}^{m_1} \, r_{yy}^{m_2} \, (2r_{xy})^{m_3} \sum_{p_{1} + \cdots +p_{N} = m_1}^{m_1} \, \sum_{q_{1} +  \cdots +q_{N} = m_2}^{m_2} \nonumber \\
        &\times \sum_{r_{1} + \cdots +r_{N} = m_3}^{m_3} \, \left[ \prod_{i=1}^N \frac{1}{p_i! \, q_i! \, r_i!} \right] \int \diff \bm a \, \left( \prod_{i=1}^N a_i^{h_i - 1} \right) \, \delta \left( \sum_{i=1}^N a_i - t_1 \right)  \nonumber \\
        &  \times \int \diff \bm b \, \left( \prod_{i=1}^N b_i^{k_i - 1} \right) \,  \delta \left( \sum_{i=1}^N b_i - t_2 \right) \left[\prod_{i=1}^N a_i^{2p_{i} + r_{i}} \, b_i^{ 2q_i + r_i  } \right] \label{eq:Phi_preLT} 
    \end{align}
    \endgroup
    and so using exactly the same steps as in Eqs. \eqref{eq:int_steps_start} to \eqref{eq:int_steps_finish} we can evaluate the integrals over $\bm a$ and $\bm b$. Inserting the result from this process into Eq. \eqref{eq:Phi_preLT}, we see that
    \begin{align}
        \Phi_{\bm x, \bm y, \bm h, \bm k}(m) =& \frac{m! }{B(\bm h) B(\bm k)} \sum_{m_1 + m_2 + m_3 = m}^m \frac{ r_{xx}^{m_1} \, r_{yy}^{m_2} \, (2r_{xy})^{m_3}}{\Gamma(2m_1 + m_3 + \eta) \, \Gamma(2m_2 + m_3 + \kappa)} \nonumber \\
        & \times \sum_{r_1=0}^{m_3} \frac{1}{r_1!} \cdots \sum_{r_N=0}^{m_3} \frac{1}{r_N!} \, \delta_{\sum_i r_i, m_3} \sum_{p_1 + \cdots + p_N = m_1}^{m_1} \left[ \prod_{i=1}^N \frac{\Gamma(2p_i + r_i + h_i)}{p_i!} \right] \nonumber \\
        & \times \sum_{q_1 + \cdots + q_N = m_2}^{m_2} \left[ \prod_{i=1}^N \frac{\Gamma(2q_i + r_i + k_i)}{q_i!} \right] \label{eq:Phi_pre_KDs}
    \end{align}
    and so we introduce and evaluate the following function 
    \begin{align}
        S_{\bm h}(m_1, \bm r) \equiv& \sum_{p_1 + \cdots  +p_N = m_1}^{m_1} \left[ \prod_{i=1}^N \frac{\Gamma(2p_i + r_i + h_i)}{p_i!} \right] \nonumber \\
        =& \frac{1}{m_1!} \lim_{w \to 0} \frac{\diff^{m_1}}{\diff w^{m_1}} \left[ \prod_{i=1}^N \left( \sum_{j=0}^{m_1} \frac{\Gamma(2j + r_i + h_i)}{j!} w^j \right) \right] \ ,
        \label{eq:S_def}
    \end{align}
    with the second equality being achieved using Eq. \eqref{eq:multi_sum_2} of Lemma \ref{lem:multinomial_sum}. This then allows us to write that
    \begin{align}
        \Phi_{\bm x, \bm y, \bm h, \bm k}(m) =\frac{m!}{B(\bm h) B(\bm k)} \sum_{m_1 + m_2 + m_3 = m}^m \frac{ r_{xx}^{m_1} \, r_{yy}^{m_2} \, (2r_{xy})^{m_3}}{\Gamma(2m_1 + m_3 + \eta) \, \Gamma(2m_2 + m_3 + \kappa)} \nonumber \\
        \times \chi_{\bm h, \bm k}(m_1, m_2, m_3) \ ,
    \end{align}
    where
    \begingroup
    \allowdisplaybreaks
    \begin{align}
        \chi_{\bm h, \bm k}(m_1, m_2, m_3) \equiv& \sum_{r_1=0}^{m_3} \frac{1}{r_1!} \cdots \sum_{r_N=0}^{m_3} \frac{1}{r_N!} \, S_{\bm h}(m_1, \bm r) \, S_{\bm k}(m_2, \bm r) \, \delta_{\sum_i r_i, m_3} \nonumber \\
        =& \frac{1}{m_3!} \lim_{w \to 0} \frac{\diff^{m_3}}{\diff w^{m_3}} \left[ \sum_{r_1=0}^{m_3} \frac{w^{r_1}}{r_1!} \cdots \sum_{r_N=0}^{m_3} \frac{w^{r_N}}{r_N!} \, S_{\bm h}(m_1, \bm r) \, S_{\bm k}(m_2, \bm r) \right] \ . \label{eq:chi_def_sums}
    \end{align}
    \endgroup
    Finally, one can perform the summation over $m_3$ using the fact that $m_1 + m_2 + m_3 = m$ but also maintaining that $m- (m_1 + m_2) > 0$. After renaming $m_1 \to i$ and $m_2 \to j$ this yields the required result of Eq. \eqref{eq:moms_tau_M2}.

    We now move onto the proof of the integer moments of the product of the squared singular values, given around Eq. \eqref{eq:moms_Delta_ab}. The product of the non-zero singular values squared can be written as $\Delta = \Delta_{xy} \Delta_{ab}$, where $\Delta_{a b} \equiv (\bm a^T \bm a) (\bm b^T \bm b) - (\bm a^T \bm b)^2$ and $\Delta_{x y} \equiv (\bm x^T \bm x) (\bm y^T \bm y) - (\bm x^T \bm y)^2$. In this set up, both $\bm x$ and $\bm y$ are fixed vectors of  length $N$, hence all disorder is contained solely in the object $\Delta_{ab}$ and so we introduce
    \begin{align}
        \Psi_{\bm h, \bm k}(m, t_1, t_2) \equiv \frac{1}{B(\bm h) B(\bm k)} &\int \diff \bm a \, \left( \prod_{i=1}^N a_i^{h_i - 1} \right) \int \diff \bm b \, \left( \prod_{i=1}^N b_i^{k_i - 1} \right) \delta \left( \sum_{i=1}^N a_i - t_1 \right) \nonumber \\
        &\times \delta \left( \sum_{i=1}^N b_i - t_2 \right) \bigg( ( \bm a^T \bm a )( \bm b^T \bm b) - (\bm a^T \bm b)^2  \bigg)^m  \ ,
    \end{align}
     with auxiliary variables $t_1$ and $t_2$ used for the sums of the general DRVs, such that $\Psi_{\bm h, \bm k}(m) \equiv \Psi_{\bm h, \bm k}(m,1,1) = \expval{\Delta_{ab}^m}_{\bm a, \bm b}$. As we have done previously, we now utilise the multinomial theorem which allows us to write that
     \begin{align}
        \Psi_{\bm h, \bm k}(m, t_1, t_2) =& \frac{m!}{B(\bm h) B(\bm k)} \sum_{m_1+m_2=m}^m \frac{(-1)^{m_2}}{m_1! \, m_2!} \int \diff \bm a \, \left( \prod_{i=1}^N a_i^{h_i - 1} \right) \int \diff \bm b \, \left( \prod_{i=1}^N b_i^{k_i - 1} \right) \nonumber \\
        &\times \delta \left( \sum_{i=1}^N a_i - t_1 \right) \delta \left( \sum_{i=1}^N b_i - t_2 \right) \, ( \bm a^T \bm a)^{m_1} ( \bm b^T \bm b)^{m_1} (\bm a^T \bm b)^{2m_2} \ ,
    \end{align}
    and doing so again on the scalar products we see that
    \begin{align}
        \Psi_{\bm h, \bm k}(m, t_1, t_2) =& \frac{m!}{B(\bm h) B(\bm k)} \sum_{m_1+m_2=m}^m \frac{(-1)^{m_2} \, m_1! \, (2m_2)!}{m_2!} \sum_{p_1 + \cdots +p_N=m_1}^{m_1} \sum_{q_1 + \cdots + q_N=m_1}^{m_1}   \nonumber \\
        &\times \sum_{r_1 + \cdots +r_N=2m_2}^{2m_2} \left[\prod_{i=1}^N \frac{1}{p_i! \, q_i! \, r_i!} \right] \int \diff \bm a \, \delta \left( \sum_{i=1}^N a_i - t_1 \right)\nonumber\\
        &\times \int \diff \bm b \, \delta \left( \sum_{i=1}^N b_i - t_2 \right) \left[ \prod_{i=1}^N a_i^{2p_i + r_i + h_i - 1} b_i^{2q_i + r_i + k_i - 1} \right] \ .
    \end{align}
    One can then evaluate the integrals using exactly the same steps as outlined in Eqs. \eqref{eq:int_steps_start} to \eqref{eq:int_steps_finish}, after some further simple manipulations it can be seen that
    \begin{align}
        \Psi_{\bm h, \bm k}(m) =& \frac{m!}{B(\bm h) B(\bm k)} \sum_{m_1+m_2=m}^m \frac{(-1)^{m_2} \, m_1! \, (2m_2)!}{m_2! \, \Gamma(2m + \eta) \, \Gamma(2m + \kappa)} \sum_{r_1=0}^{2m_2} \frac{1}{r_1!} \cdots \sum_{r_N=0}^{2m_2} \frac{1}{r_N!} \delta_{\sum\limits_i r_i, 2m_2} \nonumber \\
        & \times \sum_{p_1+ \cdots +p_N=m_1}^{m_1} \prod_{i=1}^N \frac{\Gamma(2p_i + r_i + h_i)}{p_i!} \sum_{q_1 + \cdots +q_N = m_1}^{m_1} \prod_{i=1}^N \frac{\Gamma(2q_i + r_i + k_i)}{q_i!} 
    \end{align}
    and through the use of Eq. \eqref{eq:S_def}, we can identify the second line of the above equation as $S_{\bm h}(m_1, \bm r) S_{\bm k}(m_1, \bm r)$ with $\bm r = (r_1, r_2, \ldots, r_N)^T$. Furthermore, we can use the definition of $\chi_{\bm h, \bm k}(m_1,m_2,m_3)$ given in Eq. \eqref{eq:chi_def_sums}, to write that
    \begin{align}
        \Psi_{\bm h, \bm k}(m) =& \frac{m!}{B(\bm h) B(\bm k)} \sum_{m_1+m_2=m}^m \frac{(-1)^{m_2} \, m_1! \, (2m_2)!}{m_2! \, \Gamma(2m + \eta)  \, \Gamma(2m + \kappa)} \, \chi_{\bm h, \bm k}(m_1, m_1, 2 m_2) \nonumber \\
        =& \frac{m!}{B(\bm k) B(\bm h)} \sum_{n=0}^m \frac{(-1)^{m-n} \, n! \, (2(m-n))!}{(m-n)! \, \Gamma(2m + \eta)  \, \Gamma(2m + \kappa)} \, \chi_{\bm h, \bm k}\Big(n, n, 2 (m-n) \Big) \ ,
    \end{align}
    leading to Eq. \eqref{eq:moms_Delta_ab} - thus completing our proof of the results in Theorem \ref{thm:sig2_M2}.
\end{proof}

\section{Proofs of results for rank-$R$ NNEMs} \label{sec:rankR_proofs}

In this section, we provide proofs of the results outlined in Theorems \ref{thm:lam_MR} and \ref{thm:sig_MR} related to $M_R$, which is an NNEM of rank-$R$. We start in Section \ref{sec:proofs_lam_MR} by deriving the results in Theorem \ref{thm:lam_MR} pertaining to the eigenvalues of $M_R$. Then we move onto the proof of the results related to the squared singular values in Section \ref{sec:proofs_sig_MR}.

\subsection{Non-zero eigenvalues of rank-$R$ NNEMs}
\label{sec:proofs_lam_MR}

We now provide proofs of Eqs. \eqref{eq:pdf_T_MR} and \eqref{eq:moms_T_MR} which describe the pdf and moments of the sum of the eigenvalues of the rank-$R$ NNEM $M_R$. We then move onto prove Eq. \eqref{eq:mom1_DR}, which gives the first moment of the product of the associated non-zero eigenvalues. 

\begin{proof}
    Considering the NNEM $M_R$, described in Definition \ref{def:MR}, we look to prove the result of Eq. \eqref{eq:pdf_T_MR} for the distribution of the sum of its eigenvalues, denoted as $T_R$. In this set-up
    \begin{equation}
        M_R = \sum_{i=1}^R \bm z_i \bm a_i^T \hspace{1cm} \text{and} \hspace{1cm} T_R = \sum_{i=1}^R \bm a_i^T  \bm z_i \equiv \sum_{i=1}^R \lambda_i \ ,
    \end{equation}
    where, for $i = 1,2, \ldots, R$, $\lambda_i \equiv \bm a_i^T  \bm z_i$ and $\bm z_i$ are fixed vectors with $N$ non-negative entries. Additionally, we have $R$ independent vectors of Dirichlet random variables $\bm a_i$, where the Dirichlet parameters are contained in an associated vector $\bm k_i$, with the sums of their entries denoted by $\kappa_{i} = \sum_j (\bm k_i)_j$. The independence of each $\bm a_i$ implies that the set of all $\lambda_i$ are independent variables, with their individual distributions being attainable from Eq. \eqref{eq:pdf_lambda_form1}. For clarity, we now repeat this result using the notation of our rank-$R$ matrix as
    \begin{equation}
        \mathcal{P}_{\bm z_i, \bm k_i}(\lambda_i) = \frac{\Gamma(\kappa_i)(-1)^{\kappa_i-1}}{\Gamma(\kappa_i - 1)} \bm \partial_{\bm x_i}^{\bm k_i} \left[ \sum_{j_i=1}^N \omega_{ij_i} (\lambda_i - X_{ij_i})^{\kappa_i-2}  \Theta\left[\lambda_i - X_{ij_i} \right] \right]_{\bm x_i = \bm z_i}
        \label{eq:P_TR_start}
    \end{equation}
    with
    \begin{equation}
        \omega_{ij_i} = \left(\prod\limits_{\ell \neq j_i}^N (X_{ij_i} - X_{i \ell})\right)^{-1} \ ,
    \end{equation}
    where $X$ is an $R \times N$ matrix whose entries are the necessary differential variables and are later set to $X_{ij} = Z_{ij}$. Note that, as alluded to in the proof of Theorem \ref{thm:lam_M1}, the proof of Eq. \eqref{eq:pdf_T_MR} is most easily done using the form of $\mathcal{P}_{\bm z_i, \bm k_i}(\lambda_i)$ in Eq. \eqref{eq:P_TR_start}
    and not the otherwise equivalent form corresponding to Eq. \eqref{eq:pdf_lambda_form2}. Using the definition of the pdf of each $\lambda_i$, we can see that the distribution of $T_R$ can be derived from
    \begin{align}
        \mathcal{P}&_{Z, K}(T_R) \equiv \left[\prod_{i=1}^R \int_0^\infty \diff \lambda_i \, \mathcal{P}_{\bm z_i, \bm k_i}(\lambda_i) \right] \delta\left( T_R - \sum_{i=1}^R \lambda_i \right)=  C_{K,R} \, \bm \partial_{X}^{K} \Bigg[\sum_{j_1=1}^N \omega_{1j_1} \sum_{j_2=1}^N \omega_{2j_2} \cdots \nonumber \\
        & \times \sum_{j_R=1}^N \omega_{Rj_R} \left[\prod_{i=1}^R \int_0^\infty \diff \lambda_i \left( \lambda_i - X_{ij_i} \right)^{\kappa_i-2} \Theta\left[ \lambda_i - X_{ij_i} \right]\right] \delta\left( T_R - \sum_{i=1}^K \lambda_i \right) \Bigg]_{X=Z} \ , \label{eq:P_T_rank_pre_LT}
    \end{align}
    where we have introduced
    \begin{equation}
        C_{K,R} = \prod_{i=1}^R \frac{\Gamma(\kappa_i)(-1)^{\kappa_i - 1}}{\Gamma(\kappa_i - 1)} \hspace{0.5cm} \text{and} \hspace{0.5cm} \bm \partial_{X}^K = \prod_{i=1}^R \bm \partial_{\bm x_i}^{\bm k_i} = \prod_{i=1}^R \prod_{j=1}^N \frac{1}{\Gamma(K_{ij})} \frac{\partial^{K_{ij} - 1}}{\partial X_{ij}^{K_{ij} - 1}}\ .
    \end{equation}
    For now we focus only on the terms inside the integrals and write that
    \begin{align}
        I_{X}\left(T_R\right) \equiv& \left[\prod_{i=1}^R \int_0^\infty \diff \lambda_i \left( \lambda_i - X_{ij_i} \right)^{\kappa_i-2} \Theta\left[ \lambda_i - X_{ij_i} \right]\right] \delta\left( T_R - \sum_{i=1}^R \lambda_i \right) \ ,
    \end{align}
    then we perform a Laplace transform w.r.t. $T_R$ such that
    \begin{align}
        \widetilde{I}_{X}(\tau) \equiv& \int_0^\infty \diff T_R \, e^{- \tau T_R} I_{X}(T_R) \nonumber \\
        =& \left[\prod_{i=1}^R \int_{X_{ij_i}}^\infty \diff \lambda_i \left( \lambda_i - X_{ij_i} \right)^{\kappa_i-2} e^{- \tau \lambda_i} \right] = \frac{\prod\limits_{i=1}^R \Gamma(\kappa_i - 1) }{\tau^{\widehat{\kappa}_R - R} } e^{-\tau \sum\limits_{i=1}^R X_{ij_i} } \ ,
    \end{align}
    where $\widehat{\kappa}_R = \sum_{i=1}^R \kappa_i$. The associated Laplace inversion can then be easily carried out and, after inserting the result back into Eq. \eqref{eq:P_T_rank_pre_LT}, we find that
    \begin{align}
        \mathcal{P}_{Z, K}(T_R) =& \frac{\left(\prod\limits_{i=1}^R \Gamma(\kappa_i)\right) (-1)^{\widehat{\kappa}_R - R} }{\Gamma(\widehat{\kappa}_R - R)} \bm \partial_{X}^K \Bigg[ \sum_{j_1=1}^N \omega_{1j_1} \sum_{j_2=1}^N \omega_{2j_2} \cdots \nonumber \\
        &\times \sum_{j_R=1}^N \omega_{Rj_R} \left( T_R - \sum_{i=1}^R X_{ij_i}\right)^{\widehat{\kappa}_R - R -1} \Theta\left[ T_R - \sum_{i=1}^R X_{ij_i} \right] \Bigg]_{X=Z} \ . 
    \end{align}
    Finally, we can use the result derived in \ref{app:sym_funcs_rankR} to see that
    \begin{align}
        \mathcal{P}_{Z, K}(T_R) = \frac{\left(\prod\limits_{i=1}^R \Gamma(\kappa_i)\right) }{\Gamma(\widehat{\kappa}_R - R)}& \bm \partial_{X}^K \Bigg[ \sum_{j_1=1}^N \omega_{1j_1} \sum_{j_2=1}^N \omega_{2j_2} \cdots \sum_{j_R=1}^N \omega_{Rj_R} \nonumber \\
        & \times \left( \sum_{i=1}^R X_{ij_i} - T_R \right)^{\widehat{\kappa}_R - R -1} \Theta\left[ \sum_{i=1}^R X_{ij_i} - T_R \right] \Bigg]_{X=Z} \ ,
    \end{align}
    thus concluding our proof of Eq. \eqref{eq:pdf_T_MR}. With this pdf now in place we can also easily obtain the associated non-negative integer moments ($m \geq 0$) of $T_R$. These can be derived as
    \begin{align}
        \expval{T_R^m}&_A \equiv \int_0^\infty \diff T_R \, T_R^m \, \mathcal{P}_{Z, K}(T_R) \nonumber \\
        =& \frac{\left(\prod\limits_{i=1}^R \Gamma(\kappa_i)\right) \Gamma(m+1) }{\Gamma(\widehat{\kappa}_R + m + 1 - R)} \bm \partial_{X}^K \left[ \sum_{j_1=1}^N \omega_{1j_1} \cdots \sum_{j_R=1}^N \omega_{Rj_R} \left( \sum_{i=1}^R X_{ij_i} \right)^{\widehat{\kappa}_R +m - R}  \right]_{X=Z}  \ , 
        \label{eq:moms_TR_proof}
    \end{align}
    exactly as presented in Eq. \eqref{eq:moms_T_MR}. Now that we have results for the distribution and moments of the sum of the eigenvalues of the rank-$R$ NNEM $M_R$, we can easily specialise these results to $R=2$ to obtain the associated quantities for the rank-$2$ NNEM, $M_2$ - outlined in Eqs. \eqref{eq:pdf_T_M2} and \eqref{eq:moms_T_M2}. Finally, with Eq. \eqref{eq:moms_TR_proof} in place, we can also demonstrate that this pdf is appropriately normalised by considering the zeroth moment. This proceeds starting with
    \begin{align}
        \int_0^\infty \diff T_R \, \mathcal{P}_{Z, K}(T_R) 
        =& \frac{\left(\prod\limits_{i=1}^R \Gamma(\kappa_i)\right) }{\Gamma(\widehat{\kappa}_R + 1 - R)} \bm \partial_{X}^K \left[ \sum_{j_1=1}^N \omega_{1j_1} \cdots \sum_{j_R=1}^N \omega_{Rj_R} \left( \sum_{i=1}^R X_{ij_i} \right)^{\widehat{\kappa}_R - R}  \right]_{X=Z} \ ,
    \end{align}
    which by expanding and utilising the multinomial theorem can be written as
    \begin{align}
        \int_0^\infty \diff T_R \, \mathcal{P}_{Z, K}(T_R) =& \Bigg(\prod\limits_{i=1}^R \Gamma(\kappa_i)\Bigg) \sum_{p_1 + \ldots + p_R = \widehat{\kappa}_R - R}^{\widehat{\kappa}_R - R} \frac{1}{p_1! \, p_2! \cdots p_R!} \nonumber \\
        &\times \bm \partial_{\bm x_1}^{\bm k_1} \left[ \sum_{j_1=1}^N \omega_{1j_1} X_{1j_1}^{p_1} \right]_{\bm x_1 = \bm z_1} \cdots \bm \partial_{\bm x_R}^{\bm k_R} \left[ \sum_{j_R=1}^N \omega_{Rj_R} X_{Rj_R}^{p_R} \right]_{\bm x_R = \bm z_R} \ . 
    \end{align}
    If one considers Eq. \eqref{eq:deriv_Sm}, we can see that each of the above actions of the differential operator $\bm \partial_{\bm x_i}^{\bm k_i}$, for $i=1, \ldots, R$, produces a non-zero result only for $p_i \geq \kappa_i -1$. Since the $p_i$s are constrained to sum to $\widehat{\kappa}_R-R$, we note that the only possible choice that yields a non-zero contribution is to use $p_i = \kappa_i - 1$ for $i = 1, 2, \ldots ,R$. Using the same argument as presented to derive Eq. \eqref{eq:norm_arg}, we note that, for these values of $p_i$, the action of $\bm \partial_X^K$ yields unity and so 
    \begin{equation}
        \int_0^\infty \diff T_R \, \mathcal{P}_{Z, K}(T_R) = 1 \ , 
    \end{equation}
    meaning that this pdf is appropriately normalised.
    
    Further to the distribution of the sum of the eigenvalues, in Theorem \ref{thm:lam_MR} we also give a result for the first moment of the product of the non-zero eigenvalues of $M_R$, given in Eq. \eqref{eq:mom1_DR}. Noting that $D_R = \det(\Lambda_R)$, we can use the fact that the determinant of an $R\times R$ matrix can be expressed as a sum over all permutations of the $R$-tuple $\{1, 2, \ldots, R \}$, denoted by $P_R$, such that 
    \begin{equation}
        \det(\Lambda_R) = \sum_{p \in P_R} \sgn \left( p \right) \left( \bm a_1^T \bm z_{p(1)}\right) \left( \bm a_2^T \bm z_{p(2)}\right) \cdots \left( \bm a_R^T \bm z_{p(R)}\right)
        \label{eq:det_lam_R_expanded}
    \end{equation}
    where $p$ is a particular permutation of $\{1,2, \ldots, R\}$ and $\textup{sgn}(p)$ is the signature of the permutation, i.e. it is equal to $+1 (-1)$ if $p$ is obtained by making an even (odd) number of permutations to $\{1,2, \ldots, R \}$. Noting that each vector $\bm a_i$, for $i = 1, 2, \ldots, R$, only appears once in the product for each particular permutation, we can write that 
    \begin{equation}
        \expval{\det(\Lambda_R)}_{A} = \sum_{p \in P_R} \textup{sgn} \left( p \right) \expval{\bm a_1^T \bm z_{p(1)}}_{\bm a_1} \expval{\bm a_2^T \bm z_{p(2)}}_{\bm a_2} \cdots \expval{\bm a_R^T \bm z_{p(R)}}_{\bm a_R} 
        \label{eq:exp_DR_pre}
    \end{equation}
    and so, generally, we only require the first moment of $\expval{\bm a^T \bm z}_{\bm a}$ for a general $\bm a$ and $\bm z$, where, for now, we are suppressing the use of the indexing subscripts. In Eq. \eqref{eq:moms_lambda_GDRV}, we derived an equation for all integer moments of the random object $\lambda = \bm a^T \bm z$, setting $m=1$ in this result we see that
    \begin{equation}
        \expval{\bm a^T \bm z}_{\bm a} = \frac{ 1 }{ \kappa } \, \bm \partial_{\bm x}^{\bm k} \, \left[ \sum_{j=1}^N \frac{x_j^{\kappa }}{\prod\limits_{\ell \neq j}^N (x_j - x_\ell)} \right]_{\bm x = \bm z} 
    \end{equation}
    and if we now utilise Eq. \eqref{eq:S_m} to simplify the summation and Eq. \eqref{eq:def_partials} for the definition of $\bm \partial_{\bm x}^{\bm k}$ we see that 
    \begin{align}
        \expval{\bm a^T \bm z}_{\bm a} = \frac{ 1 }{ \kappa } \left( \prod_{i=1}^N \frac{1}{\Gamma(k_i)} \frac{\partial^{k_i - 1}}{\partial x_i^{k_i - 1}} \right) \left[ \sum\limits_{i_1, \ldots , i_N = 0}^{\kappa-N+1} x_1^{i_1} \cdots x_N^{i_N} \delta_{\sum_j i_j, \kappa-N+1 } \right]_{\bm x = \bm z} \ .
    \end{align}
    Noting that for each of the derivatives to be non-zero we require $i_j \geq k_j - 1$, for $j=1, 2, \ldots , N$, we can write the above as
    \begin{align}
        \expval{\bm a^T \bm z}_{\bm a} =& \frac{ 1 }{ \kappa } \left( \prod_{i=1}^N \frac{1}{\Gamma(k_i)} \frac{\partial^{k_i - 1}}{\partial x_i^{k_i - 1}} \right) \left[ \sum\limits_{i_1=0}^{1} x_1^{\kappa_1 - 1 + i_1} \cdots \sum\limits_{i_N=0}^{1} x_N^{\kappa_N - 1 + i_N} \delta_{\sum_j i_j, 1 } \right]_{\bm x = \bm z} \nonumber \\
        =& \frac{ 1 }{ \kappa } \left( \prod_{i=1}^N \frac{1}{\Gamma(k_i)} \frac{\partial^{k_i - 1}}{\partial x_i^{k_i - 1}} \right) \left[ \sum_{j=1}^N \left(\prod_{i=1}^N x_i^{\kappa_i-1}\right) x_j \right]_{\bm x = \bm z} = \frac{\bm k^T \bm z}{\kappa} \ .
    \end{align}
    Thus, if we now return to Eq. \eqref{eq:exp_DR_pre}, it can be seen that
    \begin{equation}
        \expval{D_R}_{A} = \left( \prod_{i=1}^R \frac{1}{\kappa_i}\right) \sum_{p \in P_R} \sgn \left( p \right) \left( \bm k_1^T \bm z_{p(1)} \right) \cdots  \left( \bm k_R^T \bm z_{p(R)} \right) \ ,
    \end{equation}
    and then we recognise that the remaining summation can be expressed as $\det(KZ^T)$ - thus completing the proof of Eq. \eqref{eq:mom1_DR}. 
\end{proof}

\subsection{Non-zero singular values of rank-$R$ NNEMs}
\label{sec:proofs_sig_MR}

In Theorem \ref{thm:sig_MR} we presented the expressions for the first moment of the sum and product of the squared non-zero singular values of $M_R$, given by Eqs. \eqref{eq:mom1_tauR} and \eqref{eq:mom1_DeltaR}, respectively. Below we present proofs of these two results.

\begin{proof}
    To attain Eq. \eqref{eq:mom1_tauR}, which describes the first moment of the sum of the squared singular values of $M_R$, we start from the definition of $\tau_R$ in Eq. \eqref{eq:def_tau_MR} and note that the average w.r.t. $A$ can be written as
    \begin{equation}
        \expval{\tau_R}_{A}  = \sum_{i,j=1}^N \expval{\bm a_i^T \bm a_j}_{\bm a_i, \bm a_j} \bm z_j^T \bm z_i  \ .
    \end{equation}
    Thus, we need to attain $ \expval{\bm a_i^T \bm a_j}_{\bm a_i, \bm a_j}$ for the cases of $i = j$ and $i \neq j$. Starting with the simpler case of equal indices, $i=j$, and dropping the subscript, we can easily obtain a closed form for the average by using $m=1$ in Eq. \eqref{eq:mom_phi_chi_def}:
    \begin{align}
        \expval{\varphi}_{\bm a} \equiv& \expval{\bm a^T \bm a}_{\bm a} = \frac{1}{B(\bm k) \Gamma(\kappa + 2)} \sum_{m_1 = 0}^1 \frac{\Gamma(2m_1 + k_1)}{\Gamma(m_1 + 1)} \cdots \sum_{m_N = 0}^1 \frac{\Gamma(2m_N + k_N)}{\Gamma(m_N + 1)} \delta_{\sum_i m_i, 1} \nonumber \\
        =& \frac{1}{B(\bm k) \Gamma(\kappa + 2)} \sum_{j=1}^N \left( \prod_{i\neq j}^N \Gamma(k_i) \right) \Gamma(k_j + 2) = \frac{\bm k^T \bm k + \kappa}{\kappa(\kappa + 1)} \ .
    \end{align}
    To calculate $\expval{\bm a_i^T \bm a_j}_{\bm a_i, \bm a_j}$ for $i \neq j$ we borrow some notation from our work on rank-2 matrices and say that $\bm a_i = \bm a$ and $\bm a_j = \bm b$, with Dirichlet parameters arranged in the vectors $\bm h$ and $\bm k$, respectively. In this way we can write that
    \begin{align}
        \expval{\bm a^T \bm b}_{\bm a, \bm b} = \frac{1}{B(\bm h) B(\bm k)} \sum_{j=1}^N &\int \diff \bm a \left( \prod_{i=1}^N a_i^{h_i - 1} \right) a_j \, \delta\left( \sum_{i=1}^N a_i - 1 \right) \label{eq:aTb_pre_int} \\
        &\times \int \diff \bm b \left( \prod_{i=1}^N b_i^{k_i - 1} \right) b_j \, \delta\left( \sum_{i=1}^N b_i - 1 \right) \ . \nonumber
    \end{align}
    We can evaluate the integrals over $\bm a$ and $\bm b$ separately, using an object of the form
    \begin{equation}
        f_{\bm h}(t; j) \equiv \frac{1}{B(\bm h)} \int \diff \bm a \left( \prod_{i=1}^N a_i^{h_i - 1} \right) a_j \, \delta\left( \sum_{i=1}^N a_i - t \right) \ ,
    \end{equation}
    with auxiliary variable $t$ used to denote the sum of the DRVs. If we now perform a Laplace transform on this object we see that 
    \begin{align}
        \widetilde{f}_{\bm h}(s ; j) \equiv& \int_0^\infty dt \, e^{-st} f_{\bm h}(t; j) = \frac{1}{B(\bm h)} \int \diff \bm a \left( \prod_{i=1}^N a_i^{h_i - 1} \right) a_j \, e^{-s\sum_i a_i} \nonumber\\
        =& \frac{1}{B(\bm h)} \left( \prod_{i \neq j}^N \frac{\Gamma(h_i)}{s^{h_i}} \right) \frac{\Gamma(h_j+1)}{s^{h_j+1}} = \Gamma(\eta) \frac{h_j}{s^{\eta + 1}} \ ,
    \end{align}
    then, after the associated Laplace inversion, we can return to Eq. \eqref{eq:aTb_pre_int} and see that 
    \begin{equation}
        \expval{\bm a^T \bm b}_{\bm a, \bm b} = \frac{1}{\eta \kappa} \sum_{j=1}^N h_j k_j = \frac{\bm h^T \bm k}{\eta \kappa} \ .
    \end{equation}
    Thus, restoring the index notation, we can see that
    \begin{equation}
        \expval{\bm a_i^T \bm a_j}_{\bm a_i, \bm a_j} = 
        \begin{cases}
            \frac{\bm k_i^T \bm k_i + \kappa_i}{ \kappa_i (\kappa_i + 1)} \hspace{1cm}&[i = j] \\
            \frac{\bm k_i^T \bm k_j}{ \kappa_i \kappa_j} \hspace{1cm}&[i \neq j]
        \end{cases} \ ,
    \end{equation}
    which verifies Eq. \eqref{eq:mom1_tauR}. Finally, we now discuss the result given in Eq. \eqref{eq:mom1_DeltaR} and in particular look to evaluate the expectation value,
    \begin{equation}
       \expval{\det(\Sigma_A)}_{A} = \sum_{p \in P_R} \sgn(p) \expval{\bm a_1^T \bm a_{p(1)} \bm a_2^T \bm a_{p(2)} \cdots \bm a_R^T \bm a_{p(R)} }_A \ ,
      \label{eq:mom1_SigmaA}
    \end{equation}
    where $p$, $P_R$ and $\sgn(p)$ are defined below Eq. \eqref{eq:det_lam_R_expanded}. The main technical issue in evaluating this object comes in calculating the expectation of a `loop' of $s$ inner products of Dirichlet random vectors. To explain the procedure we discuss it in detail for $R=3$, making clear how analogous steps can be performed at higher $R$. At $R=3$ we can see that
    \begin{align}
        \expval{\det(\Sigma_A)}_A =& \Big\langle \bm a_1^T \bm a_1 \, \bm a_2^T \bm a_2 \, \bm a_3^T \bm a_3 - \bm a_1^T \bm a_1 \, \bm a_2^T \bm a_3 \, \bm a_3^T \bm a_2 - \bm a_2^T \bm a_2 \, \bm a_1^T \bm a_3 \, \bm a_3^T \bm a_1 \nonumber \\
        & \hspace{0.4cm} - \bm a_3^T \bm a_3 \, \bm a_1^T \bm a_2 \, \bm a_2^T \bm a_1 + \bm a_1^T \bm a_2 \, \bm a_2^T \bm a_3 \, \bm a_3^T \bm a_1 + \bm a_1^T \bm a_3 \, \bm a_3^T \bm a_2 \, \bm a_2^T \bm a_1 \Big\rangle_{\bm a_1, \bm a_2, \bm a_3} \nonumber \\
        =& \expval{ \bm a_1^T \bm a_1 }_{\bm a_1} \expval{ \bm a_2^T \bm a_2}_{\bm a_2} \expval{  \bm a_3^T \bm a_3}_{\bm a_3} - \expval{\bm a_1^T \bm a_1}_{\bm a_1} \expval{\bm a_2^T \bm a_3 \, \bm a_3^T \bm a_2}_{\bm a_2, \bm a_3}  \nonumber \\
        &- \expval{\bm a_2^T \bm a_2}_{\bm a_2} \expval{\bm a_1^T \bm a_3 \, \bm a_3^T \bm a_1}_{\bm a_1, \bm a_3} - \expval{\bm a_3^T \bm a_3}_{\bm a_3} \expval{\bm a_1^T \bm a_2 \, \bm a_2^T \bm a_1}_{\bm a_1, \bm a_2} \nonumber \\
        &+ \expval{\bm a_1^T \bm a_2 \, \bm a_2^T \bm a_3 \, \bm a_3^T \bm a_1}_{\bm a_1 , \bm a_2, \bm a_3} + \expval{ \bm a_1^T \bm a_3 \, \bm a_3^T \bm a_2 \, \bm a_2^T \bm a_1 }_{\bm a_1 , \bm a_2, \bm a_3} \ ,
    \end{align}
    i.e. the result is a combination of inner products, where in some cases we can use the independence of the DRVs to perform averages separately. In general, we wish to calculate the expectation value
    \begin{equation}
         F_{K}\left( \{ \ell_1, \ell_2, \ldots , \ell_s \} \right) \equiv \expval{(\bm a_{\ell_1}^T \bm a_{\ell_2}) (\bm a_{\ell_2}^T \bm a_{\ell_3}) \cdots (\bm a_{\ell_s}^T \bm a_{\ell_1})}_{\bm a_{\ell_1}, \ldots , \bm a_{\ell_s}}
    \end{equation}
    for an arbitrary set of $1\leq s \leq R$ vectors indexed by $\{\ell_1, \ell_2 ,\ldots , \ell_s \}$. We observe that the product we are studying is a scalar, therefore it can be written as a trace and we can then utilise the cyclicity of the trace to see that
    \begin{align}
        F_{K}\left( \{ \ell_1, \ell_2, \ldots , \ell_s \} \right) =& \bigg\langle\Tr\Big( \bm a_{\ell_1} \bm a_{\ell_1}^T \, \bm a_{\ell_2} \bm a_{\ell_2}^T  \cdots \bm a_{\ell_s} \bm a_{\ell_s}^T \Big)\bigg\rangle_{\bm a_{\ell_1}, \ldots , \bm a_{\ell_s}} 
        = \Tr\left( \prod_{i=1}^s \Psi_{\ell_i}  \right) \ .
    \end{align}
    The problem is now reduced to calculating the entries of the matrices $\Psi_{\nu} \equiv \expval{\bm a_{\nu} \bm a_{\nu}^T}_{\bm a_{\nu}}$ for $\nu = 1, 2, \ldots, R$. Considering a general Dirichlet random vector $\bm a$, with Dirichlet parameters in the vector $\bm k$, the $(i,j)$th entry of the matrix $\expval{\bm a \bm a^T}_{\bm a}$ is $\expval{a_i a_j}_{\bm a}$. Therefore, we need to calculate $\expval{a_i^2}_{\bm a}$ and $\expval{a_i a_j}_{\bm a}$ for $i,j=1, 2, \ldots, N$. These can be evaluated as
    \begin{equation}
        \expval{a_i^2}_{\bm a} = \frac{1}{B(\bm k)} \int \diff \bm a \left( \prod_{\ell =1}^N a_\ell^{k_\ell -1} \right) \delta\left( \sum_{\ell=1}^N a_\ell - 1 \right) \, a_i^2 = \frac{k_i (k_i+1)}{\kappa(\kappa + 1)}
    \end{equation}
    and 
    \begin{equation}
        \expval{a_i a_j}_{\bm a} = \frac{1}{B(\bm k)} \int \diff \bm a \left( \prod_{\ell =1}^N a_\ell^{k_\ell -1} \right) \delta\left( \sum_{\ell=1}^N a_\ell - 1 \right) \, a_i \, a_j = \frac{k_i k_j}{\kappa(\kappa + 1)} \ ,
    \end{equation}
    by employing the standard techniques utilised throughout this manuscript. With these results in place we can now write that
    \begin{equation}
        \Big( \Psi_\nu \Big)_{ij} = \frac{1}{\kappa_{\nu}(\kappa_{\nu} + 1)}\begin{cases}
            K_{\nu i} (K_{\nu i} + 1) \hspace{2cm}& [i=j] \\
            K_{\nu i} K_{\nu j} \hspace{2cm}& [i\neq j]
        \end{cases} \ \, 
    \end{equation}  
    for $\nu = 1,2, \ldots, R$, and thus returning to our example at $R=3$, we see that 
    \begin{align}
        \expval{\det(\Sigma_A)}_A =& F_K(\{1\}) F_K(\{2\}) F_K(\{3\}) - F_K(\{1\}) F_K(\{2,3\}) - F_K(2) F_K(\{1,3\}) \nonumber \\
        & - F_K(\{3\}) F_K(\{1,2\}) + F_K(\{ 1 , 2 ,3 \}) + F_K(\{ 1 , 3, 2 \}) \ .
    \end{align}
    As noted previously, one can easily extend this procedure to cases of higher ranks.
\end{proof}

\subsection*{Acknowledgements}
Y.V.F. acknowledges financial support from EPSRC Grant EP/V002473/1 ``Random Hessians and Jacobians: theory and applications''. M.J.C. is supported by a studentship from the Faculty of Natural and Mathematical Sciences at King's College London. 

\appendix
\let\oldsection\thesection
\renewcommand{\thesection}{\oldsection}
\addtocontents{toc}{\setlength{\cftsecnumwidth}{15ex}}

\section{Identity for rank-$R$ NNEMs}
\label{app:sym_funcs_rankR}

In the derivation of the pdf of $T_R$ for rank-$R$ matrices, the following combination of symmetric functions arises after we have performed the inverse Laplace transform
\begin{align}
    S^{(1)}_{K, R}(T;X) \equiv (-1)^{\widehat{\kappa}_R - R} \bm \partial_{X}^K \left[ \sum_{j_1=1}^{N} \omega_{1 j_1} \cdots \sum_{j_R=1}^{N} \omega_{R j_R} \left( T - \sum_{i=1}^R X_{i j_i} \right)^{\widehat{\kappa}_R - R - 1} \Theta\left[ T - \sum_{i=1}^R X_{i j_i} \right]  \right]
\end{align}
and we now wish to show that this is equivalent to the following different combination
\begin{align}
    S^{(2)}_{K, R}(T;X) \equiv \bm \partial_{X}^K \left[ \sum_{j_1=1}^{N} \omega_{1 j_1} \cdots \sum_{j_R=1}^{N} \omega_{R j_R} \left( \sum_{i=1}^R X_{i j_i} - T \right)^{\widehat{\kappa}_R - R - 1} \Theta\left[ \sum_{i=1}^R X_{i j_i} - T \right]  \right] \ .
\end{align}
We start with choosing a value of $T = T_1$ such that for all combinations of $\{ j_1, j_2, \ldots, j_R \}$, $T_1 > \sum_{i=1}^{R} X_{ij_i}$. Immediately, we see that $S^{(2)}_{K, R}(T_1;X) = 0$ and furthermore
\begin{align}
    S^{(1)}_{K, R}(T_1;X) =& (-1)^{\widehat{\kappa}_R - R} \bm \partial_{X}^K \left[ \sum_{j_1=1}^{N} \omega_{1 j_1} \cdots \sum_{j_R=1}^{N} \omega_{R j_R} \left( T_1 - \sum_{i=1}^R X_{i j_i} \right)^{\widehat{\kappa}_R - R - 1} \right] \nonumber \\
    =& (-1)^{\widehat{\kappa}_R - R} \sum_{\underset{\widehat{\kappa}_R-R-1}{p_1 + \cdots + p_R + \tilde{p}_T=} }^{\widehat{\kappa}_R - R -1} \frac{(\widehat{\kappa}_R - R - 1)!}{p_1! \cdots p_R! \, \tilde{p}_T!} \, T_1^{\tilde{p}_T} (-1)^{p_1 + \cdots +p_R} \, \nonumber  \\
    &\hspace{2cm} \times \bm \partial_{\bm x_1}^{\bm k_1} \left( \sum_{j_1 = 1}^N \omega_{1 j_1} X_{1j_1}^{p_1}\right) \cdots \, \bm \partial_{\bm x_R}^{\bm k_R} \left( \sum_{j_R = 1}^N \omega_{R j_R} X_{Rj_R}^{p_R}\right) \ . \label{eq:S_T1}
\end{align}  
In order to consider the product on the final row of the above equation, we treat the first term and employ the findings to the other terms. Using Eq. \eqref{eq:S_m}, we can see that
\begin{equation}
    \bm \partial_{\bm x_1}^{\bm k_1} \left( \sum_{j_1 = 1}^N \omega_{1 j_1} X_{1j_1}^{p_1}\right) = \bm \partial_{\bm x_1}^{\bm k_1} \left( \sum_{j_1 = 1}^N \frac{ X_{1j_1}^{p_1}  }{\prod_{\ell \neq j_1}(X_{1j_1} - X_{1 \ell} )  } \right) = \begin{cases}
        0 & \hspace{0.5cm} [p_1 = 0, 1, \ldots , \kappa_1 - 2] \\
        > 0 & \hspace{0.5cm} [p_1 = \kappa_1 - 1, \kappa_1, \ldots ] \\
    \end{cases}
\end{equation}
i.e. this contribution is only non-zero when $p_1 \geq \kappa_1 - 1$. Bearing in mind that for Eq. \eqref{eq:S_T1} to be non-zero, we require that each $p_i \geq \kappa_i - 1$ for $i = 1, 2, \ldots , R$, the sum $p_1 + p_2 + \cdots + p_R \geq \widehat{\kappa}_R - R$, this imposes that $S^{(1)}_{K, R}(T_1;X)=0$ as the sum of the $p_i$s is at most equal to $\widehat{\kappa}_R - R - 1$. Furthermore, taking a value of $T=T_2$, such that for all $\{ j_1, j_2, \ldots, j_R\}$, $T_2 < \sum_{i=1}^R X_{ij_i}$ we can follow a similar method to what has been outlined above and find that $S^{(1)}_{K, R}(T_2;X) = S^{(2)}_{K, R}(T_2;X) = 0$. 

For the final part of this proof we must choose a value of $T = T_3$ such that $T_3 > \sum_{i=1}^{R}X_{ij_i}$ for some subset $J_T$ of $J$, where $J$ is the set of all $\{j_1, j_2, \ldots, j_R\}$ for $j_i = 1,\ldots, N$ and $i=1,\ldots,R$. In this case we can employ a similar argument to what we have done previously, so as to see that
\begin{align}
    (-1)^{\widehat{\kappa}_R - R} \bm \partial_{X}^K \left[ \sum_{j_1=1}^{N} \omega_{1 j_1} \cdots \sum_{j_R=1}^{N} \omega_{R j_R} \left( T_3 - \sum_{i=1}^R X_{i j_i} \right)^{\widehat{\kappa}_R - R - 1} \Theta\left[ T_3 - \sum_{i=1}^R X_{i j_i} \right]  \right] = 0
\end{align}
which allows us to write that
\begin{align}
    S_{K, R}^{(1)}(T_3; &X) = (-1)^{\widehat{\kappa}_R - R}
	 \bm \partial_{X}^K \left[ \sum_{j_1, \ldots, j_R \in J_T }^{N} \omega_{1 j_1} \cdots \omega_{R j_R} \left( T_3 - \sum_{i=1}^R X_{i j_i} \right)^{\widehat{\kappa}_R - R - 1}  \right] \nonumber \\
     =& - (-1)^{\widehat{\kappa}_R - R} \bm \partial_{X}^K \left[ \sum_{j_1, \ldots, j_R \in J \backslash J_T }^{N} \omega_{1 j_1} \cdots \omega_{R j_R} \left( T_3 - \sum_{i=1}^R X_{i j_i} \right)^{\widehat{\kappa}_R - R - 1}  \right] \nonumber \\
     =& \bm \partial_{X}^K \left[ \sum_{j_1, \ldots, j_R \in J \backslash J_T }^{N} \omega_{1 j_1} \cdots \omega_{R j_R} \left(  \sum_{i=1}^R X_{i j_i} - T_3 \right)^{\widehat{\kappa}_R - R - 1}  \right] = S_{K, R}^{(2)}(T_3; X) \ .
\end{align}
Therefore, we have proved that for an arbitrary $R\times N$ matrix $X$ and for any $T \in R$, we have that
\begin{equation}
    S_{K, R}^{(1)}(T; X) = S_{K, R}^{(2)}(T; X) \ .
    \label{eq:app_sym_funcs_results}
\end{equation}

\end{document}